%% file: main.tex
\documentclass{article}

\usepackage[utf8]{inputenc} % allow utf-8 input
\usepackage[T1]{fontenc}    % use 8-bit T1 fonts
\usepackage{hyperref}       % hyperlinks
\usepackage{url}            % simple URL typesetting
\usepackage{booktabs}       % professional-quality tables
\usepackage{amsfonts}       % blackboard math symbols
\usepackage{nicefrac}       % compact symbols for 1/2, etc.
\usepackage{microtype}      % microtypography
\usepackage{xcolor}         % colors

\usepackage{listings}
\usepackage{minted}
\usepackage{tcolorbox}
\tcbuselibrary{minted,skins, breakable}
\usepackage{mdframed}
\usepackage{fancyvrb}
\usepackage{wrapfig}
\usepackage{tcolorbox}
\usepackage{mdframed}
\usepackage{multicol}
\usepackage{microtype}
\usepackage{tikz}
\usetikzlibrary{automata, positioning, arrows.meta, shapes.geometric}
\usepackage{graphicx}
\usepackage{subfigure}
\usepackage{algorithm}
\usepackage{float}
\usepackage{algpseudocode}
\usepackage{multirow}
\usepackage{colortbl}
\usepackage{algpseudocode}
\usepackage{booktabs} % for professional tables
\usepackage{multirow}
\usepackage{multicol}
\usepackage{anyfontsize}
\usepackage{xcolor}
\usepackage{pifont}   
\usepackage{threeparttable}
\usepackage{xspace}
\usepackage{stfloats}
\usepackage{makecell}
\usepackage{subcaption}
\definecolor{Green}{RGB}{0,128,0}
\definecolor{lightgray}{RGB}{225,225,23}
\usepackage{hyperref}

\newcommand{\name}{\texttt{SymRTLO}\xspace}
\newcommand{\eg}{\textit{e.g., }}
\newcommand{\ie}{\textit{i.e., }}
\usepackage[final]{neurips_2025}

\usepackage{amsmath}
\usepackage{amssymb}
\usepackage{mathtools}
\usepackage{amsthm}
\usepackage{graphicx}
\usepackage{titlesec}
\newcommand*\circled[1]{\tikz[baseline=(char.base)]{
    \node[shape=circle,draw,inner sep=1pt] (char) {#1};}}
\usepackage[capitalize,noabbrev]{cleveref}

\theoremstyle{plain}

\theoremstyle{definition}

\theoremstyle{remark}

\usepackage[textsize=tiny]{todonotes}

\title{SymRTLO: Enhancing RTL Code Optimization with LLMs and Neuron-Inspired Symbolic Reasoning}

\author{%
  \textbf{Yiting Wang}$^{1,*}$ \quad
  \textbf{Wanghao Ye}$^{1,*}$ \quad
  \textbf{Ping Guo}$^{2,*}$ \quad
  \textbf{Yexiao He}$^{1,*}$ \quad
  \textbf{Ziyao Wang}$^{1}$ \\
  \textbf{Bowei Tian}$^{1}$ \quad
  \textbf{Shwai He}$^{1}$ \quad
  \textbf{Guoheng Sun}$^{1}$ \quad
  \textbf{Zheyu Shen}$^{1}$ \quad
  \textbf{Sihan Chen}$^{3}$ \\
  \textbf{Ankur Srivastava}$^{1}$ \quad
  \textbf{Qingfu Zhang}$^{2}$ \quad
  \textbf{Gang Qu}$^{1}$ \quad
  \textbf{Ang Li}$^{1}$ \\
  \\
  \\
  $^{1}$University of Maryland, College Park \quad
  $^{2}$City University of Hong Kong \quad
  $^{3}$University of Southern California \\
  $^{*}$Equal contribution \quad
 $^{\dagger}$Corresponding author: angli@cs.umd.edu
}

\begin{document}

\maketitle

\begin{abstract}

Optimizing Register Transfer Level (RTL) code is crucial for improving the efficiency and performance of digital circuits in the early stages of synthesis. Manual rewriting, guided by synthesis feedback, can yield high-quality results but is time-consuming and error-prone. Most existing compiler-based approaches have difficulty handling complex design constraints. Large Language Model (LLM)-based methods have emerged as a promising alternative to address these challenges. However, LLM-based approaches often face difficulties in ensuring alignment between the generated code and the provided prompts. This paper introduces SymRTLO, a neuron-symbolic framework that integrates LLMs with symbolic reasoning for the efficient and effective optimization of RTL code. Our method incorporates a retrieval-augmented system of optimization rules and Abstract Syntax Tree (AST)-based templates, enabling LLM-based rewriting that maintains syntactic correctness while minimizing undesired circuit behaviors. A symbolic module is proposed for analyzing and optimizing finite state machine (FSM) logic, allowing fine-grained state merging and partial specification handling beyond the scope of pattern-based compilers. Furthermore, a fast verification pipeline, combining formal equivalence checks with test-driven validation, further reduces the complexity of verification. Experiments on the RTL-Rewriter benchmark with Synopsys Design Compiler and Yosys show that SymRTLO improves power, performance, and area (PPA) by up to 43.9\%, 62.5\%, and 51.1\%, respectively, compared to the state-of-the-art methods.

\end{abstract}

\input{sections/intro}
\input{sections/background2}

\input{sections/methodology}

\input{sections/experiments}

\input{sections/conclusion}

\nocite{langley00}

\bibliography{main}
\bibliographystyle{plain}

%%%%%%%%%%%%%%%%%%%%%%%%%%%%%%%%%%%%%%%%%%%%%%%%%%%%%%%%%%%%%%%%%%%%%%%%%%%%%%%
%%%%%%%%%%%%%%%%%%%%%%%%%%%%%%%%%%%%%%%%%%%%%%%%%%%%%%%%%%%%%%%%%%%%%%%%%%%%%%%
% APPENDIX
%%%%%%%%%%%%%%%%%%%%%%%%%%%%%%%%%%%%%%%%%%%%%%%%%%%%%%%%%%%%%%%%%%%%%%%%%%%%%%%
%%%%%%%%%%%%%%%%%%%%%%%%%%%%%%%%%%%%%%%%%%%%%%%%%%%%%%%%%%%%%%%%%%%%%%%%%%%%%%%
\newpage
\appendix
\onecolumn
\input{sections/appendix}

\end{document}

%% file: sections/intro.tex
\section{Introduction}

% 1. optimization of rtl is important 2.currently heavy rely on human rewriting: 1. needs expertise, time consuming, 2. complexity of design can be hard for manual rewriting. 3.Currently the compiler based method: 1. has issue with complex design optimization, 2. does not show optimized code in rtl level, cannot optimize via synthesis feedback. 

Register Transfer Level (RTL) optimization is a cornerstone of modern circuit design flows, serving as the foundation for achieving optimal Power, Performance, and Area (PPA). As the earliest phase in the hardware design lifecycle, RTL development provides engineers with the most significant degree of flexibility to explore design patterns, make architectural trade-offs, and influence the overall design quality~\cite{chu2006rtl}. Engineers use hardware description languages (HDLs) like Verilog to describe circuit functionality. At this stage, decisions made have far-reaching implications, as the quality of the RTL implementation directly impacts subsequent stages, including synthesis, placement, and routing ~\cite{wang2009electronic}. A well-optimized RTL not only ensures better design outcomes but also prevents suboptimal designs from propagating through the flow, leading to significant inefficiencies and costly iterations \cite{10323951, 994596}.

Despite its importance, RTL optimization remains a challenging and labor-intensive task. Engineers must iteratively refine their designs through multiple rounds of synthesis and layout feedback to ensure functionality and meet stringent PPA targets. This process becomes increasingly cumbersome as design complexity grows, with synthesis times scaling disproportionately, often taking hours or even days for a single iteration.\cite{10323951} Consequently, designers frequently face numerous synthesis cycles to evaluate trade-offs and reach acceptable results. While modern electronic design automation (EDA) tools provide compiler-based methods to aid optimization, these approaches are inherently limited.~\cite{rtlrewriter} They rely heavily on predefined heuristics, making them ill-suited for adapting to unconventional design patterns, complex constraints, or dynamic optimization scenarios. As a result, the RTL optimization process demands significant expertise and effort.

%edit here : focus on two challenges 1.Not Forcing alignment between RAG provided and the outcome 2. Not optimization goal focused 

Recent advances in artificial intelligence, particularly the advent of large language models (LLMs), have introduced a new paradigm for automating and optimizing RTL code. Leveraging the powerful generative capabilities of LLMs, researchers have demonstrated their potential to rewrite and optimize Verilog code automatically~\cite{rtlrewriter}. However, existing LLM-based approaches face critical challenges that limit their effectiveness. First, these models often fail to align their generated outputs with specified optimization objectives. The inherent limitations in logical reasoning within LLMs can lead to deviations from intended transformations, resulting in suboptimal or incorrect outputs. Second, despite their potential for automating code generation, current methods still heavily rely on traditional synthesis feedback loops for optimization. This reliance results in the inefficiencies of the synthesis process, failing to address the core issue of long design cycles.

%\paragraph{Proposed Framework.}
%To address these gaps, we propose \textbf{SymRTLOptimizer}, a \emph{neuron-symbolic} RTL optimization framework that integrates LLM-driven rewriting with symbolic reasoning to achieve both automation and explainability. The core components of our approach are:

%modify here: what problem to solve
%\begin{itemize}
   % \item \textbf{Template-Based Optimization Module:}  
   % A library of AST-based templates ensures syntactic correctness for straightforward transformations (e.g., for datapaths or multiplexers). This reduces LLM-related errors by providing safe, predefined patterns that the model can adapt.

   % \item \textbf{Optimization Rule and Goal-Based Search Engine:}  
   % We encode domain expertise in a rule-based retrieval-augmented system, where each rule targets a specific optimization goal (area, power, or timing). This guides the LLM and avoids the pitfalls of purely pattern-based generation.

    %\item \textbf{Symbolic System Construction:}  
   % By segmenting circuits and focusing on deep semantic extraction (especially for FSM logic), our framework builds symbolic models of the code. These models permit case-by-case optimizations that go beyond generic AST rewrites and handle partial or long-context specifications more robustly.

   % \item \textbf{Verification System:}  
   % We implement a fast verification pipeline that automates testbench generation, solver selection (depending on combinational or sequential logic), and equivalence checking. This reduces the verification cost and ensures that only functionally correct rewrites are accepted.
%\end{itemize}

\vspace{-1em}

\paragraph{Our Proposed Framework.} 

To address the critical challenges in RTL optimization, we introduce \name, the first neuron-symbolic system that seamlessly integrates LLMs with symbolic reasoning to optimize RTL code. \name significantly reduces the reliance on repeated synthesis tool invocations and enhances the alignment of LLM-generated results with intended optimization rules. 

%The overall workflow of \name is illustrated in Figure~\ref{fig:ARCH_compx}.

% \begin{figure*}[ht]
%     \centering
%     \includegraphics[width=\textwidth]{SymbolicLLM_ARCH.pdf}
%     \caption{SymRTLOPtimizer Architecture}
%     \label{fig:ARCH_compx}
% \end{figure*}

Designing such a system introduces significant challenges, which we address through the integration of carefully designed modules. The first challenge lies in the \textit{generalization} of optimization rules. Traditionally, optimization patterns are scattered across code samples, books, and informal notes, making it difficult for compiler-based methods to formalize or apply them effectively. \name tackles this by employing an LLM-based rule extraction system, combined with a retrieval-augmented generation (RAG) mechanism and a search engine. This ensures that optimization rules are not only generalized but also efficiently retrievable from a robust library built from diverse sources.

Another critical challenge is the \textit{alignment} of LLM-generated RTL code with the intended transformations, as LLMs often struggle to produce outputs that strictly adhere to the specified optimization objective, leading to unreliable and unexplainable results. To ensure alignment, \name employs Abstract Syntax Tree (AST)-based templates, which guide the LLM to generate code that satisfies syntactic and semantic correctness. For complex control flows or edge cases that exceed the capabilities of AST templates, the framework utilizes a symbolic generation module, designed to handle such scenarios dynamically while maintaining optimization quality.

In addition to alignment, \textit{conflicts} often arise when different design patterns are required to meet distinct PPA  goals. To address this, \name introduces a goal-oriented approach, where each optimization rule is explicitly tied to its intended objective. This enables selective application based on user-defined optimization goals,  efficiently balancing these conflicts to deliver optimized designs without disproportionately compromising PPA metrics.

\textit{Verification} in traditional RTL workflows demands significant manual effort for test case development. To address this, \name integrates an automated test case generator, streamlining verification while ensuring functional correctness.

Our key contributions are summarized as follows:
\vspace{-1em}
\begin{itemize}
    \item \textbf{LLM Symbolic Optimization:} \name, the first framework to combine LLM-based rewriting with symbolic reasoning for RTL optimization.
    \vspace{-1em}
    \item \textbf{Data Path and Control Path Optimization:} \name addresses critical challenges in both traditional EDA compilers and purely LLM-based approaches, particularly by aligning generated code with FSM and data path algorithms, balancing conflicting optimization rules, and improving explainability.
    \vspace{-1em}
    \item \textbf{PPA Improvements:} \name demonstrates its efficacy on industrial-scale and benchmark circuits, surpassing manual coding, classical compiler flows (\eg Synopsys DC Compiler), and state-of-the-art LLM-based methods, achieving up to 43.9\%, 62.5\%, and 51.1\% improvements in power, delay, and area.
    \vspace{-1em}
\end{itemize}

%% file: sections/background2.tex
\vspace{10pt}
\section{Background and Motivation}

\vspace{-5pt}

LLMs have emerged as powerful tools for RTL design automation, with various approaches being developed since 2023. As summarized in Table~\ref{tab:llmcompare}, these approaches fall into three primary categories: \textit{RTL code generation}~\cite{liu:2023:chipnemo,liu:2024:rtlcoder,chang:2023:chipgpt,blocklove:2023:chipchat}, \textit{debugging}~\cite{tsai:2024:rtlfixer}, and \textit{optimization}~\cite{rtlrewriter}. This growing body of research demonstrates the significant potential of LLMs to improve the efficiency and effectiveness of EDA workflows. However, RTL code optimization has always been a significant challenge in RTL design, even for human experts, as it has the greatest impact on the performance of downstream tasks.

%RTL rewriter -LLM: consistency, long synthesis time
\textbf{Challenges in RTL Code Optimization with LLMs.} Aligning generated code with intended optimization goals is a major challenge in LLM-based RTL optimization. Due to inherent randomness, LLMs often produce incomplete, incorrect, or suboptimal results. For example, RTLRewriter~\cite{rtlrewriter} employs retrieval-augmented prompts and iterative synthesis-feedback loops to enhance functional correctness but still struggles with fundamental misalignment between generated code and optimization objectives. Additionally, the need for multiple synthesis rounds significantly increases optimization time as design complexity increases, limiting the scalability of current LLM-based methods for large industrial-scale designs.
\definecolor{green}{RGB}{0,153,0}
\definecolor{red}{RGB}{170,0,0}
\begin{wraptable}{r}{0.55\textwidth}
    \setlength{\tabcolsep}{1.5pt}
    \centering
    \caption{Comparative analysis of LLM-based methods for RTL design. \textcolor{green}{\ding{52}} indicates the presence of the feature, while \textcolor{red}{\ding{56}} indicates absence of the feature.}
    \label{tab:llmcompare}
    \fontsize{7}{10}\selectfont  % Change the second parameter to control line spacing
    \renewcommand{\arraystretch}{0.75}  % This will now work without resizebox
    \begin{tabular}{l|l|cccc}
        \toprule
        \textbf{Category} & \textbf{Method} & \textbf{Verification} & \textbf{Rule}& \textbf{Output} & \textbf{Conflict} \\
        & & \textbf{Capability}& \textbf{-Based}& \textbf{Alignment} & \textbf{Resolution} \\
        \midrule
        \multirow{6}{*}{\textbf{Generation}} 
            & ChipNeMo~\cite{liu:2023:chipnemo} & \textcolor{red}{\ding{56}} & \textcolor{red}{\ding{56}} & \textcolor{red}{\ding{56}} & \textcolor{red}{\ding{56}} \\
            & VeriGen~\cite{thakur:2024:verigen} & \textcolor{red}{\ding{56}} & \textcolor{red}{\ding{56}} & \textcolor{red}{\ding{56}} & \textcolor{red}{\ding{56}} \\
            & VerilogEval~\cite{liu:2023:verilogeval} & \textcolor{red}{\ding{56}} & \textcolor{red}{\ding{56}} & \textcolor{red}{\ding{56}} & \textcolor{red}{\ding{56}} \\
            & RTLCoder~\cite{liu:2024:rtlcoder} & \textcolor{red}{\ding{56}} & \textcolor{red}{\ding{56}} & \textcolor{red}{\ding{56}} & \textcolor{red}{\ding{56}} \\
            & ChipChat~\cite{blocklove:2023:chipchat} & \textcolor{red}{\ding{56}} & \textcolor{red}{\ding{56}}& \textcolor{red}{\ding{56}} & \textcolor{red}{\ding{56}} \\
            & ChipGPT~\cite{chang:2023:chipgpt} & \textcolor{red}{\ding{56}} & \textcolor{red}{\ding{56}}& \textcolor{red}{\ding{56}} & \textcolor{red}{\ding{56}} \\
        \midrule
        \textbf{Debug} & RTLFixer~\cite{tsai:2024:rtlfixer} & \textcolor{green}{\ding{52}}& \textcolor{green}{\ding{52}}& \textcolor{red}{\ding{56}}& \textcolor{red}{\ding{56}}\\\midrule
        \multirow{2}{*}{\textbf{Optimization} }& RTLRewriter~\cite{rtlrewriter} & \textcolor{green}{\ding{52}}& \textcolor{green}{\ding{52}}& \textcolor{red}{\ding{56}}& \textcolor{red}{\ding{56}}\\
    & \textbf{\name (Ours)}& \textcolor{green}{\ding{52}}& \textcolor{green}{\ding{52}}& \textcolor{green}{\ding{52}}& \textcolor{green}{\ding{52}}\\
        \bottomrule
    \end{tabular}
\end{wraptable}

%Engineers ways : 1. underuse of documents - optimization generalization rules3. verificaiton done by human

\vspace{-3pt}
\textbf{Current Approaches: Underutilization of Knowledge and Manual Verification.} Traditional RTL design optimization relies on established patterns such as subexpression elimination~\cite{subexpression, cocke1970global}, dead code elimination~\cite{knoop1994partial, gupta1997path}, strength reduction~\cite{cooper2001operator}, algebraic simplification~\cite{buchberger1982algebraic, carette2004understanding}, Mux reduction~\cite{chen2004register, wang2023muxoptimization}, and memory sharing~\cite{laforest10, ma2020hypervisor}. While these techniques are effective, these optimizations typically operate at the gate-level netlist, making the relationship between optimized output and original RTL code less transparent. Additionally, optimization patterns from design manuals and codebases remain underutilized due to the lack of a centralized repository, forcing engineers to rely on their expertise rather than automated tools. Furthermore, verification requires the creation of test benches and test cases manually, making the RTL design flow both time-consuming and error-prone.

\textbf{Optimization Conflicts and Limited Compiler Capabilities.} Existing compiler-based methods face additional challenges, particularly in managing \textbf{optimization goal conflicts} and handling complex patterns. For instance, optimizing for one metric, such as delay, often conflicts with another, such as power consumption. Striking the right balance between these competing objectives is crucial, especially as trade-offs between power and delay directly impact overall system performance. As shown in Table~\ref{table:ppa_conflicts}, each optimization method has its own specific goal, which often clashes with others. Compiler-based methods also lack the flexibility to adapt to such conflicts, limiting their effectiveness in optimizing designs with diverse and competing constraints.

\vspace{-5pt}
\begin{table}[h]
\renewcommand{\arraystretch}{1.1} 
\fontsize{7}{8}\selectfont
\centering

\vspace{-5pt}
\begin{minipage}{0.48\textwidth}
\centering
\setlength{\tabcolsep}{1pt}
\caption{Conflicts Between Design Goals \& Optimization Patterns.}

\begin{tabular}{l|c|c|c}
\hline
\textbf{Design Pattern } & \textbf{Goal} & \textbf{Conflict Goal} & \textbf{Conflict Design Pattern} \\ \hline
Pipelining & Low Timing & Low Area & Resource Sharing \\ 
Clock Gating & Low Power & Low Timing & Retiming \\ \hline
\end{tabular}
\label{table:ppa_conflicts}
\end{minipage}
\hfill
\begin{minipage}{0.48\textwidth}
\centering
\setlength{\tabcolsep}{2pt}
\caption{Performance Comparison Across Different Approaches.}
\begin{tabular}{l|c|c|c|c}
\hline
\textbf{Approach} & \textbf{\# States} & \textbf{Time (ns)} & \textbf{Power (mW)} & \textbf{Area ($\mu$m$^2$)} \\ \hline
Baseline & 11 & 0.041 & 2.250 & 833.000\\
GPT-O1 & 10 & 0.041 & 2.280 & 993.480 \\
Optimized & 4 & 0.025 & 1.170 & 403.920\\\hline
\end{tabular}
\label{tab:comparison}
\end{minipage}
\end{table}
\vspace{1pt}% 2. conflicts 

% For Control Flow oprimization methods FSM state minimization~\cite{kam2013synthesis} and state assignment techniques~\cite{villa2012synthesis}. have been 

\textbf{Advancing LLM-based RTL Optimization with Neuro-Symbolic Integration.} Recent research has shown a growing trend toward combining symbolic reasoning with LLM \cite{yang2023neurosymbolic, wan2024neurosymbolicTowards, chen2024pyod2, calanzone2024neurosymbolic, tilwani2024neurosymbolic, wan2024neurosymbolic}, bringing new inspirations for more efficient and reliable LLM-based RTL code optimization with better prompt-code alignment. These integrated approaches have seen broad application across various fields, from automated theorem proof and knowledge representation to robotics and medical diagnostics, demonstrating how the combination of  pattern recognition and generative capabilities of LLM with the interpretability and logical rigor of symbolic systems can significantly improve the alignment between LLM output and the given prompt.

\textbf{Motivation Experiments.} To highlight the limitations of current LLM-based RTL optimization, we conduct an experiment using state-of-the-art commercial LLM, GPT-O1, to optimize an 11-state FSM design. The goal was to minimize and merge unnecessary states to enhance PPA metrics. GPT-O1 receives a detailed state reduction algorithm to guide the optimization process. We compare its results with an optimized design that directly applied the state reduction algorithm. As shown in Table~\ref{tab:comparison}, GPT-O1 struggles to align its outputs with the algorithm, resulting in an under-optimized FSM with minimal state reduction and improvement in PPA. In contrast, algorithm-driven optimization achieves significantly better results, highlighting current LLM limitations in complex RTL optimization tasks.

%% file: sections/methodology.tex
\vspace{-3pt}
\section{Methodology}
\vspace{-3pt}

\begin{figure*}[t]
    \centering
    \includegraphics[width=1\textwidth]{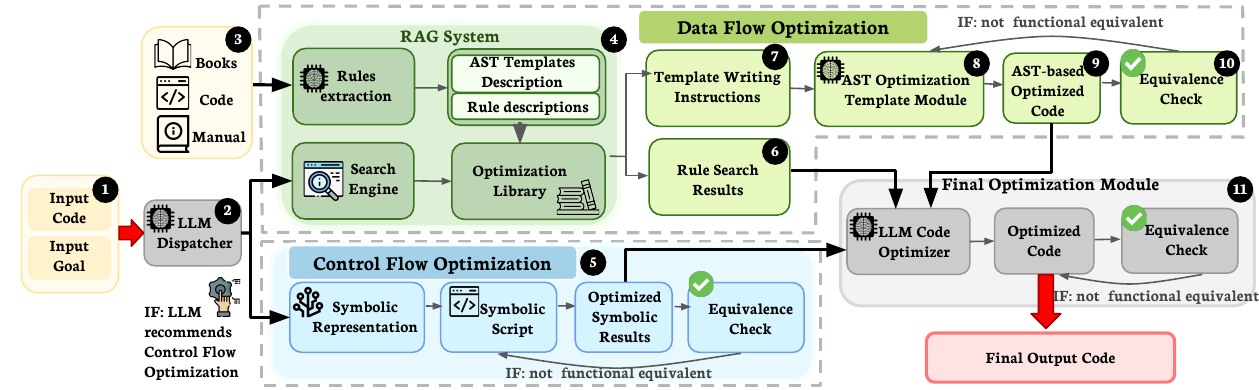}
    \caption{SymRTLO Architecture.}
    \label{fig:ARCH_compx}
\end{figure*}

\name takes a Verilog RTL module as input and optimizes it for specific design goals, such as low power, high performance, or reduced area. As illustrated in Figure~\ref{fig:ARCH_compx}, the workflow begins by entering the RTL code and the user-specified optimization goal (\circled{1}) into the \textbf{LLM Dispatcher} (\circled{2}). This dispatcher analyzes the input circuit and determines the appropriate optimization path: either proceeding solely with \textbf{Data Flow Optimization} (\S\ref{dataflow}) or incorporating \textbf{Control Flow Optimization} (\S\ref{subsec:control-flow}) as well, depending on the characteristics of the design.  For Data Flow Optimization, a search engine with a retrieval-augmented module extracts optimization rules and constructs AST-based templates. For Control Flow Optimization, an LLM-driven symbolic system generator performs FSM-specific transformations. Finally, the \textbf{Final Optimization Module} (\S\ref{final}) integrates both paths and incorporates a verification system to ensure the functional correctness of the optimized design.

\subsection{LLM Dispatcher}
\vspace{-3pt}
The \textbf{LLM Dispatcher} (\circled{2}) receives the input RTL code and the specified optimization goal (\eg low power) (\circled{1}) before any optimization begins. It first summarizes the code and generates potential optimization suggestions. These suggestions are then passed to the Retrieval-Augmented Generation (RAG) system to identify the relevant optimization rules. Additionally, the Dispatcher evaluates the presence of a Finite State Machine (FSM) in the original code to determine whether control flow optimization is necessary.

\subsection{Data Flow Optimization}
\label{dataflow}
%alignment, generalization, conflict of optimization goal
Data Flow represents the process by which information is propagated, processed, and optimized within an RTL design. Effective data flow optimizations improve system efficiency by simplifying computations, reducing redundancy, and enhancing PPA metrics. Common techniques include sub-expression elimination, constant folding, and resource sharing.  The proposed Data Flow Optimization Module addresses three key challenges: (1) generalizing diverse optimization patterns into accurate, reusable rules; (2) aligning LLM-generated optimizations with functional and logical requirements; and (3) resolving conflicts between optimization goals inherent in distinct design patterns.
\vspace{-7pt}
%generalization, goal based, alignment
\paragraph{Optimization Rule Search Engine} 
Optimization knowledge is often scattered across books, lecture notes, codebases, and design manuals, with no generalized repository to serve as a unified knowledge base. Furthermore, optimization patterns frequently \textit{conflict} due to divergent goals, for example, power reduction versus performance improvement. To tackle these challenges, we developed a retrieval-augmented generation (RAG) system equipped with an optimization goal-based rule extraction module.
\vspace{-7pt}
%alignment
\paragraph{Optimization Library.} The RAG system aggregates raw RTL optimization data from sources such as lecture notes, manuals, and example designs~\cite{vahid2010digital, taraate2022digital, schultz2023optimizing, palnitkar2003verilog} into a comprehensive knowledge base (\circled{3}). LLMs then summarize and structure these data into an optimization library (\circled{4}). Each rule is abstracted to include its description, applicable optimization goals (\eg area, power, or timing), and its category (\eg data flow, FSM, MUX, memory, or clock gating).  A similarity engine identifies overlaps with existing entries, prompting merges or exclusive labels to ensure the scalability of the rule library. To align optimization goals with generated outputs, the rules specify detailed instructions for constructing AST templates, enabling precise application of optimization patterns. Rules with clearly defined requirements include template-writing guidelines, while more abstract rules are stored as descriptive text and used directly as optimization prompts. See Appendix~\ref{app:rag} for an example RAG. 

%Conflicts in Design Patterns dispatcher: 1.summaze code with optimization suggestions easier search, code search not easy becasue search engine 2. input optimaziton goal when giving optimization suggestion and into search engine.
\vspace{-7pt}

\paragraph{Enforcing Rule Alignment and Resolving Conflicts}
To improve the structure and alignment of optimization rules, the LLM Dispatcher (\circled{1}) provides both a summary of the input RTL code and suggestions for potential optimizations. These inputs are passed to the search engine along with user-specified optimization goals, performing a similarity search to identify the most relevant rules from the RAG system.

Given the potential for conflicts between optimization goals, it is critical to prevent the inclusion of conflicting rules while ensuring no critical optimizations are overlooked. To achieve this balance, we employ the elbow method to analyze the similarity scores between the query and the candidate rules. This approach identifies a natural cutoff point where adding more rules no longer yields significant benefits. Let the similarity scores between the query and candidate rules be ordered as: $s_1 \geq s_2 \geq \cdots \geq s_M,$
where $s_i$ denotes the similarity score for  $i$-th rule, ordered from highest to lowest, and $M$ is the total number of candidates.

The optimal cutoff index $i^*$ is determined by maximizing the difference between consecutive similarity scores: $i^* = \underset{1 \leq i < M}{\arg\max} , (s_i - s_{i+1}).$

Rules with similarity scores above the threshold $\tau_\text{elbow}$ are selected for application.
The similarity between the query embedding ($\mathbf{e}_\text{query}$) and the a rule embedding ($\mathbf{e}_\text{rule}$) is computed as shown below:
\vspace{-5pt}
\begin{equation}
\text{sim}(\mathbf{e}_\text{query}, \mathbf{e}_\text{rule}) = \frac{\mathbf{e}_\text{query} \cdot \mathbf{e}_\text{rule}}{|\mathbf{e}_\text{query}| |\mathbf{e}_\text{rule}|} \geq \tau_\text{elbow}.
\end{equation}
\vspace{-9pt}

This method ensures that only the most relevant rules are selected, striking an optimal balance between comprehensiveness and precision. The output of the search engine contains two components: rules with detailed template-writing instructions (\circled{7}) and abstract rules described only by their optimization properties (\circled{6}).

%Consistency Between LLM-Generated Code and Intended Transformations
\paragraph{AST Template Building}
To ensure that LLM-generated RTL code aligns with functional and logical optimization goals, we enforce rules using AST-based symbolic systems, which have been proven to be effective in hardware debugging~\cite{tsai:2024:rtlfixer}. Compared to LLM-generated symbolic systems, AST-based templates offer several advantages: (1) parsing Verilog into an AST ensures accurate and structured design representations; (2) limiting each template to a single optimization goal maintains conciseness, facilitating correct generation and application by LLMs; and (3) the modular approach allows selection of templates to balance conflicting optimization patterns, enhancing flexibility. 

For rules that include template-writing instructions, we prompt the LLM to generate an AST-based template that serves as a general optimization framework (\circled{8}). Let \(\mathcal{A}\) denote the set of all AST nodes in the Verilog design, and let the matching condition:
$\Phi : \mathcal{A} \rightarrow \{\text{true}, \text{false}\},$ determine whether a node qualifies for optimization. 
The process begins by identifying the \textbf{Target Node Type}, such as \texttt{Always}, \texttt{Instance}, \texttt{Assign} or \texttt{Module Instantiation}. For each node of the specified type, we apply $\Phi$ to decide whether it requires rewriting. Once the target nodes are identified, the \textbf{Transformation Rule} is applied as follows: $\tau : \{\,a \in \mathcal{A} \mid \Phi(a) = \text{true} \} \longrightarrow \mathcal{A},$
where $\tau$ replaces the matched node with an optimized AST subtree (\eg merging nested \texttt{if-else} statements, folding constants, or simplifying expressions; code example see Appendix~\ref{app:ast}). To ensure functional correctness, the transformed design undergoes an equivalence check using LLM-generated testbenches. If the template passes verification, it is stored in the RAG system as reusable content.

\begin{wrapfigure}{r}{0.5\textwidth}
    \centering
    \includegraphics[width=1\linewidth]{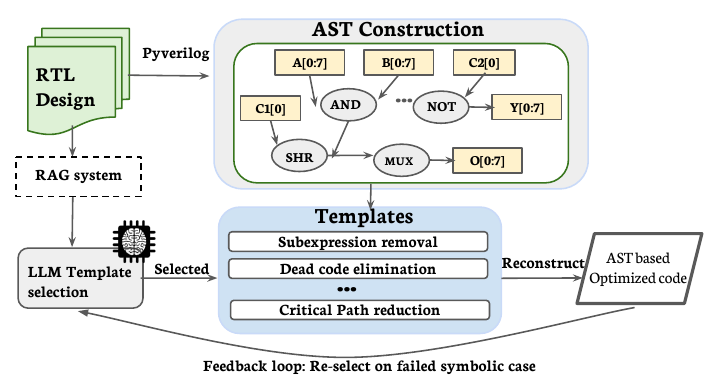}
    \caption{\name AST template optimization workflow.}
    \label{fig:template}
\end{wrapfigure}

As shown in Figure \ref{fig:template}, the RTL design is initially interpreted as an AST representation. The RAG system provides the LLM with multiple template options. Due to the varying optimization goals and scenarios, the system avoids relying on a fixed sequence of templates. Instead, the LLM determines which templates to apply and in what order, tailoring the optimization process to the design’s specific requirements. To further prevent conflicts between templates or failures in the symbolic system, we introduce a \textbf{feedback loop}. This loop allows the LLM to re-select templates and adjust its strategy based on prior failures, ensuring robustness and adaptability.

%Consistency Between LLM-Generated Code and Intended Transformations

\subsection{Control Flow Optimization}
\label{subsec:control-flow}

Control Flow, unlike Data Flow’s focus on how information is processed and propagated, defines the execution paths and sequencing of operations in RTL designs through finite-state machines (FSMs) that capture states, transitions, and outputs. These FSMs are tightly coupled with design constraints (\ie partial specifications, clock gating, and reset logic), making generic symbolic systems fragile or incomplete. Addressing these challenges requires deeper semantic analysis beyond simple pattern matching or generic AST templates. To enhance alignment between optimized code and the FSM minimization algorithm, we propose a Control Flow Optimization module utilizing an LLM-based symbolic system. An FSM can be formally represented as:
$M = (Q, \Sigma, \delta, q_0, F)$,
where $Q$ is the finite set of states, $\Sigma$ is the input alphabet, $\delta: Q \times \Sigma \rightarrow Q$ is the transition function, $q_0 \in Q$ is the initial state, and $F \subseteq Q$ is the set of accepting states.

For a partially specified FSM $M_p$, the transition function is extended to handle non-deterministic transitions:
$\delta_p: Q \times \Sigma \rightarrow 2^Q,$
where $2^Q$ represents the power set of $Q$.

Classical minimization algorithms (\eg Hopcroft’s \cite{hopcroft1969formal} or Moore’s \cite{moore1956gedanken}) are effective for fully specified deterministic FSMs but are limited by real-world complexities. Practical RTL designs often integrate control logic with data path constraints, and undefined states and transitions make FSM minimization an NP-complete problem with a general complexity of $O(2^{|Q|})$. A single pre-built AST script cannot efficiently handle all such partial specifications. Let $\phi: Q \times D \rightarrow B$ represent the data path constraints, where $D$ is the data path state space and $B$ is the boolean domain. Pure FSM-focused AST-based optimization scripts can overlook these data path side effects, failing to capture deeper control semantics.

\begin{wrapfigure}{r}{0.5\textwidth}
    \centering

    \includegraphics[width=1.1\linewidth]{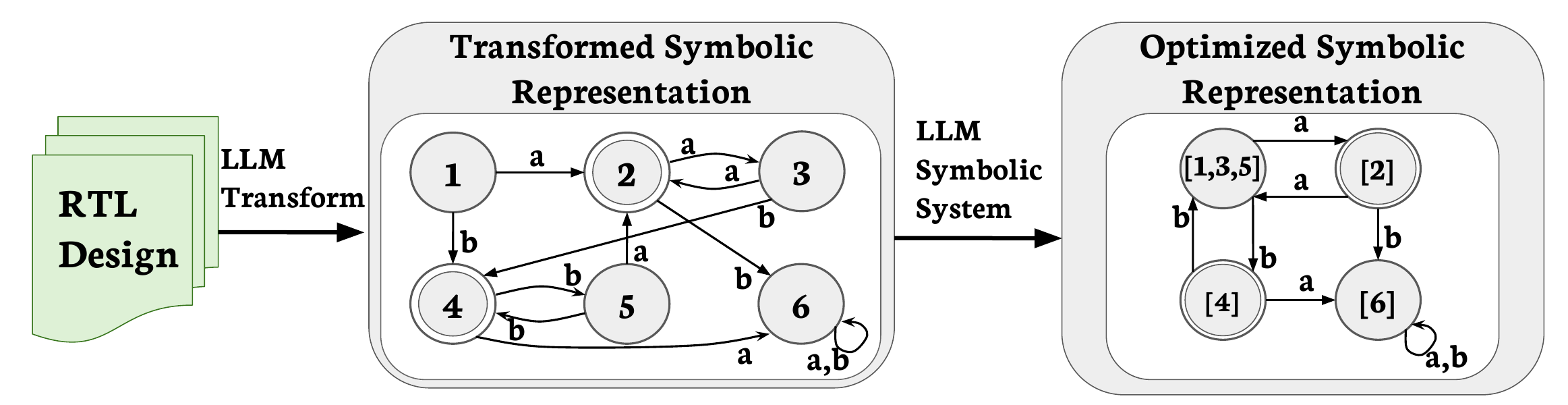}
    \caption{\name FSM optimization workflow.}
    \label{fig:fsm_sym}
\end{wrapfigure}

Inspired by \cite{hu2023codepromptingneuralsymbolic}, we propose leveraging LLMs to transform each circuit into a symbolic representation focused solely on FSM components, \ie isolating states, transitions, and relevant outputs, as illustrated in Figure \ref{fig:fsm_sym}. Instead of relying on a one-size-fits-all script, we prompt the LLM to dynamically generate a \textbf{specialized minimization script} tailored to the specific FSM structure and constraints.  We show an example workflow in Appendix~\ref{app:fsm}.

% The optimization is formulated as:\begin{equation}
% \min_{M'} f(M', C) \text{ subject to } \phi(M', D) = \text{true},
% \end{equation}
% where $M{\prime}$ represents the optimized FSM, $C$ represents the design constraints, and $\phi(M{\prime}, D)$ ensures consistency with the data path.

\vspace{-5pt}
\subsection{Verification and Final Optimization}
\vspace{-5pt}
\label{final}

To address the challenges of manual verification and unreliable automated methods, we introduce an automated verification module that integrates functionality testing and formal equivalence checking. After AST-based optimization and the LLM-assisted symbolic system generate initial results, the LLM combines extracted rules, template-optimized code, and symbolic outputs to produce the final optimized RTL design (\circled{11}). Verification is essential to ensure correctness, as LLM-generated rewrites may introduce unintended behavioral deviations. We employ a two-step verification pipeline: (1) the LLM generates test benches to validate basic functional correctness, acting as a rapid filter to reject invalid rewrites early; and (2) for designs that pass initial tests, we perform formal equivalence checking to formally confirm functional equivalence to the original design.  For combinational logic and straightforward synchronous sequential circuits, we apply standard Boolean Satisfiability Problem (SAT)-based equivalence checking. While more complex designs like asynchronous resets, CDC paths, or retimed logic, we employ advanced sequential equivalence checking with robust state matching and transformation-tolerant verification techniques.

%% file: sections/experiments.tex
\vspace{-7pt}
\section{Experiments}
\vspace{-7pt}
\subsection{Experimental Setup}

%\subsubsection{Benchmarks}

% In this sIn this study, we introduce two benchmarks for RTL
% code optimization, designed to address both long and short code sce - rios, crafted by experienced Verilog engineers. These benchmarks
% derive their optimization patterns from a comprehensive review
% of inter - l industry documents and a survey of approximately 50
% scholarly articles on RTL optimization.
\vspace{-5pt}
\paragraph{Baseline}
We compare \name with several state-of-the-art LLM and open source RTL optimization frameworks. The LLM baselines include GPT-O1, GPT-4o, GPT-4\cite{openai2023gpt4}, GPT-3.5\cite{ye2023comprehensivecapabilityanalysisgpt3}, and GPT-4o-mini. Additionally, we include two specialized open-source LLM-based tools: Verigen\cite{thakur:2024:verigen} and RTL-Coder-Deepseek \cite{liu:2024:rtlcoder}.
For a comprehensive evaluation, we analyze \name performance using circuits from the RTLRewriter Benchmark. Although the RTLRewriter environment is reproducible using Yosys \cite{wolf2013yosys} for wires and cells analysis, the exact test case they use is not provided. Moreover, comparing PPA results is even more challenging due to its reliance on Yosys + ABC \cite{BraytonMishchenko2010abc}with unknown libraries. To demonstrate \name's capabilities, we subject it to a broader evaluation scope, selecting examples that are diverse in size and functionality, while including cases reported in RTLRewriter's benchmark for direct comparison.

\vspace{-5pt}
\paragraph{Implementations}
% The \name framework uses GPT-4o as its primary LLM for tasks such as optimization strategy selection, symbolic system generation, and iterative synthesis of final HDL code. GPT-4o was chosen for its superior inference and coding capabilities, which are critical for optimization tasks. Other models like GPT-3.5 and GPT-O1 were excluded after preliminary experiments revealed poor performance: GPT-3.5 lacked sufficient coding capabilities, and GPT-O1 required significant interface enhancements between \name's modules.

The \name framework takes GPT-4o as its primary LLM for optimization strategy selection, symbolic system generation, and iterative HDL synthesis, leveraging its robust inference, low inference cost, and coding capabilities. While the framework is model-agnostic by design, alternatives like GPT-3.5 and GPT-O1 were excluded after preliminary tests showed poor results—GPT-3.5 lacked sufficient coding capability, and GPT-O1 incurred high latency and API costs, reducing overall efficiency. Pyverilog~\cite{Takamaeda:2015:ARC:Pyverilog} is used for AST extraction and code reconstruction.

To efficiently retrieve relevant transformation templates and knowledge, we integrate OpenAI's text embedding-3-small, which excels in embedding-based retrieval tasks. 
For hardware compilation and validation, we use a combination of open-source and commercial tools. Yosys measures wires and cells, while Synopsys DC Compiler 2019~\cite{synopsysDCUltra}, paired with the Synopsys Standard Cell (SSC) library, performs PPA analysis. GPT-4o generates test benches for functional coverage. Yosys + ABC serves as the logical equivalence checker,  while Synopsys Formality for sequential equivalence checking. For a fair comparison with standard compiler workflows, we apply typical Synopsys DC Compiler optimizations, using medium mapping effort and incremental mapping to reflect common practices.
\vspace{-6pt}
\paragraph{Evaluation Metrics} First, to evaluate generation quality and functional correctness, we use the pass@k metric commonly employed in code generation tasks. This metric captures the probability that at least one valid solution exists within the top $k$ generations: $\text{pass@k} = \frac{1}{N}\sum_{i=0}^N (1-\frac{C_{n_i-c_i}^k}{C_{n_i}^k})$, where $N$ is the number of problems, $n_i$ and $c_i$ represent the total and correct samples for the  $i$-th problem, respectively. Second, to test the performance of the synthesis results, we use the best results of the 10 valid generations of each model and our method. This sampling maintains token budget fairness, as SymRTLO with GPT-4o averages 7,728 tokens across all circuits, while the base GPT-4o model uses approximately 775 tokens per generation—a 10× difference.

For smaller benchmarks, we evaluate optimization results using Wires and Cells metrics, which reflect low-level physical characteristics of circuits. These metrics provide granular insights into routing complexity (wires) and logical component count (cells), offering a precise evaluation for isolated modules or blocks. For larger designs, we focus on PPA metrics to capture high-level efficiency and real-world applicability. These metrics offer a holistic view of resource usage and performance for complex designs, where low-level metrics like Wires and Cells become impractical.

\vspace{-10pt}
\subsection{Functional Correctness Analysis}

\setlength{\floatsep}{0pt} % Space between floats
\setlength{\textfloatsep}{0pt} % Space between floats and text
\setlength{\intextsep}{0pt} % Space between in-text floats and text
\vspace{-10pt}
\setcounter{table}{3} 
\renewcommand{\arraystretch}{0.5} 
\begin{wraptable}{r}{0.45\textwidth}  % Added position (r) and width parameters
    \centering
    \caption{Pass Rate Results.}
    \begingroup  % Start a group to localize the font change
    \fontsize{7}{9}\selectfont  % Adjusted line spacing for better readability
    \begin{tabular}{lccc}  % Fixed: removed one 'c' to match the number of columns
        \toprule
        \textbf{Method} & \textbf{Pass@1} &\textbf{ Pass@5} &\textbf{ Pass@10} \\
        \midrule
        Ours & \textbf{97.5} & \textbf{100.0} & \textbf{100.0} \\
        GPT-4o & 45.9 & 60.0 & 72.7 \\
        GPT-4-Turbo & 42.9 & 62.7 & 81.8 \\
        GPT-4o-mini & 2.5 & 10.9 & 12.7 \\
        GPT-3.5-Turbo & 28.6 & 42.7 & 54.5 \\
        RTL-Coder DeepSeek & 8.8 & 18.2 & 27.3 \\
        Verigen-2B & 0.0 & 0.0 & 0.0 \\
        Verigen-16B & 0.0 & 0.0 & 0.0 \\
        \bottomrule
    \end{tabular}
    \endgroup  % End the group
    \label{tab:pass_rate}
\end{wraptable}

%shorter synthesis time, alignment

To demonstrate that our method reduces synthesis time and improves functional correctness, Table \ref{tab:pass_rate} presents the evaluation results. \name achieves near-perfect first-attempt pass rates, ensuring valid, optimized RTL code with maintained functional equivalence. This significantly outperforms state-of-the-art language models, particularly given the complexity of RTL optimization tasks and the necessity of maintaining functional equivalence. \name reduces synthesis iterations, minimizing redundant computations throughout the optimization process. 

\vspace{-10pt}

\setcounter{table}{4} 
\begin{table*}[t]
\centering
\caption{Comparison of Wire and Cell Counts. Gray highlighting denotes state-of-the-art results. GeoMean is the geometric mean of the resource usage (wires or cells). Ratios are calculated by dividing the GeoMean by the baseline’s resource usage. $^\dagger$: Reported Results from \cite{rtlrewriter}.}
\label{tab:model-comparison}
\setlength{\tabcolsep}{2.4pt} 
\resizebox{1\textwidth}{!}{
{\fontsize{8}{6}\selectfont
\renewcommand{\arraystretch}{1.5} 
\begin{tabular}{@{}lcc|cc|cc|cc|cc|cc|cc|cc@{}}
\Xhline{2.5\arrayrulewidth}
\multirow{2}{*}{\textbf{Benchmark}} & \multicolumn{2}{c|}{\textbf{Yosys}} & \multicolumn{2}{c|}{\textbf{GPT-4-Turbo}} & \multicolumn{2}{c|}{\textbf{GPT-4o}} & \multicolumn{2}{c|}{\textbf{GPT-3.5-Turbo}} & \multicolumn{2}{c|}{\textbf{GPT-4o-mini}}& \multicolumn{2}{c|}{\textbf{RTLCoder-DS}} & \multicolumn{2}{c|}{\textbf{RTLrewriter}} & \multicolumn{2}{c}{\textbf{\name}} \\
& \textbf{Wires} & \textbf{Cells} & \textbf{Wires} & \textbf{Cells} & \textbf{Wires} & \textbf{Cells} & \textbf{Wires} & \textbf{Cells}  & \textbf{Wires} &\textbf{Cells}  & \textbf{Wires} & \textbf{Cells} & \textbf{Wires} & \textbf{Cells} & \textbf{Wires} & \textbf{Cells} \\
\Xhline{2.5\arrayrulewidth}
adder\_subexpression & 8 & 3 & \cellcolor{lightgray}\textbf{7} & 3 & \cellcolor{lightgray}\textbf{7} & 3 & \cellcolor{lightgray}\textbf{7} & 3  & \cellcolor{lightgray}\textbf{7}&3& 8 & 3 & \cellcolor{lightgray}\textbf{7} & 3 & \cellcolor{lightgray}\textbf{7} & 3 \\
adder\_architecture & 86 & 56 & 30 & 40 & 30 & 40 & 96 & 63  & 86&56& 86 & 56 & - & - & \cellcolor{lightgray}\textbf{14} & \cellcolor{lightgray}\textbf{16} \\
multiplier\_subexpr & 26 & 71 & \cellcolor{lightgray}\textbf{18} & \cellcolor{lightgray}\textbf{15} & 259 & 255 & 26& 71& 26&71& 26 & 71 & - & - & \cellcolor{lightgray}\textbf{18} & \cellcolor{lightgray}\textbf{15} \\
constant\_folding\_raw & 12 & 6 & 10 & 5 & 10 & 5 & 10 & 5  & 10&5& 12 & 6 & - & - & \cellcolor{lightgray}\textbf{8} & \cellcolor{lightgray}\textbf{5} \\
subexpression\_elim & 17 & 12 & 19 & 12 & 19 & 12 & 19 & 12  & 17&10& 17 & 12 & - & - & \cellcolor{lightgray}\textbf{14} & \cellcolor{lightgray}\textbf{8} \\
alu\_subexpression & 30 & 24 & 30 & 24 & 28 & 22 & 30 & 24  & 27&22& 30 & 24 & \cellcolor{lightgray}\textbf{21} & \cellcolor{lightgray}\textbf{18} & \cellcolor{lightgray}\textbf{21} & \cellcolor{lightgray}\textbf{18} \\
adder\_resource & 13 & 3 & \cellcolor{lightgray}\textbf{6} & 3 & 9 & 4 & \cellcolor{lightgray}\textbf{6} & 3  & 7&3& 13 & 3 & - & - & \cellcolor{lightgray}\textbf{6} & 3 \\
multiplier\_bitwidth & 9 & 3 & \cellcolor{lightgray}\textbf{8} & 3 & \cellcolor{lightgray}\textbf{8} & 3 & 9 & 3  & 9&3& 9 & 3 & \cellcolor{lightgray}\textbf{8} & 3 & \cellcolor{lightgray}\textbf{8} & 3 \\
multiplier\_architect & 4 & 2 & 14 & 36 & \cellcolor{lightgray}\textbf{4} & \cellcolor{lightgray}\textbf{2} & 16 & 20  & 18&36& 4 & 2 & - & - & \cellcolor{lightgray}\textbf{4} & \cellcolor{lightgray}\textbf{2} \\
adder\_bit\_width & 4 & 1 & \cellcolor{lightgray}\textbf{3} & 1 & \cellcolor{lightgray}\textbf{3} & 1 & \cellcolor{lightgray}\textbf{3} & 1  & \cellcolor{lightgray}\textbf{3}&1& 4 & 1 & \cellcolor{lightgray}\textbf{3} & 1 & \cellcolor{lightgray}\textbf{3} & 1 \\
loop\_tiling\_raw & 5 & 16 & 4 & 16 & 4 & 16 & 4 & 16  & 484&496& 5 & 16 & - & - & \cellcolor{lightgray}\textbf{3} & 16 \\
\Xhline{2.5\arrayrulewidth}
\textbf{GeoMean} & 15.49& 8.96& 13.45& 9.81& 16.35& 9.97 & 16.40& 11.49 & 16.31 & 11.49 & 15.49& 8.96& - & - & \cellcolor{lightgray}\textbf{9.75} & \cellcolor{lightgray}\textbf{5.95} \\
\textbf{Ratio} & 1.00 & 1.00 & 0.87& 1.10& 1.06& 1.11& 1.06 & 1.28 &1.11&1.31& 1.15& 0.91& 0.69$^\dagger$\footnotemark[1] & 0.77$^\dagger$\footnotemark[1] & \cellcolor{lightgray}\textbf{0.63}& \cellcolor{lightgray}\textbf{0.67}\\
\Xhline{2.5\arrayrulewidth}
\end{tabular}
}
}
\end{table*}
%\footnote[1]{$^\dagger$: Reported Results from \cite{rtlrewriter}.}

\vspace{-3pt}
\subsection{Circuit Optimization Performance}

\vspace{-5pt}
\definecolor{green}{RGB}{0,128,0}

\begin{table*}[]
\caption{FSM Designs PPA Comparison. Gray highlights indicate state-of-the-art results. A $\textcolor{green}{\boldsymbol{\Downarrow}}$ marks improvement, while a $\textcolor{red}{\boldsymbol{\Uparrow}}$ denotes a decline compare with the original design. Two comparison scenarios are shown: without compiler optimization (upper improvement) and with compiler optimization (lower improvement). A - indicates that no code is available for analysis.}

\label{tab:fsm-comparison}
\setlength{\tabcolsep}{1pt} 
\centering
\resizebox{1\textwidth}{!}{
\fontsize{8}{5}\selectfont
\renewcommand{\arraystretch}{1.8} 
\begin{tabular}{l|ccc|ccc|ccc|ccc|ccc}
\Xhline{2.5\arrayrulewidth} 
\textbf{Model/Method} & \multicolumn{3}{c|}{\textbf{example1\_state}} & \multicolumn{3}{c|}{\textbf{example2\_state}} & \multicolumn{3}{c|}{\textbf{example3\_state}} & \multicolumn{3}{c|}{\textbf{example4\_state}} & \multicolumn{3}{c}{\textbf{example5\_state}} \\
& \begin{tabular}[c]{@{}c@{}}Power\\(mW)\end{tabular} & \begin{tabular}[c]{@{}c@{}}Time\\(ns)\end{tabular} & \begin{tabular}[c]{@{}c@{}}Area\\($\mu\text{m}^2$)\end{tabular} & 
\begin{tabular}[c]{@{}c@{}}Power\\(mW)\end{tabular} & \begin{tabular}[c]{@{}c@{}}Time\\(ns)\end{tabular} & \begin{tabular}[c]{@{}c@{}}Area\\($\mu\text{m}^2$)\end{tabular} & 
\begin{tabular}[c]{@{}c@{}}Power\\(mW)\end{tabular} & \begin{tabular}[c]{@{}c@{}}Time\\(ns)\end{tabular} & \begin{tabular}[c]{@{}c@{}}Area\\($\mu\text{m}^2$)\end{tabular} & 
\begin{tabular}[c]{@{}c@{}}Power\\(mW)\end{tabular} & \begin{tabular}[c]{@{}c@{}}Time\\(ns)\end{tabular} & \begin{tabular}[c]{@{}c@{}}Area\\($\mu\text{m}^2$)\end{tabular} & 
\begin{tabular}[c]{@{}c@{}}Power\\(mW)\end{tabular} & \begin{tabular}[c]{@{}c@{}}Time\\(ns)\end{tabular} & \begin{tabular}[c]{@{}c@{}}Area\\($\mu\text{m}^2$)\end{tabular} \\
\hline
Original & 0.042 & \cellcolor{lightgray}\textbf{1.21} & 833.0 & 0.056 & 2.25 & 549.4 & 0.052 & 1.35 & 589.6 & 0.055 & 2.18 & 597.1 & 0.055 & 2.18 & 597.1 \\
GPT-3.5 & 0.043 & 1.27 & 870.6 & 0.056 & 2.25 & 549.4 & 0.052 & 1.35 & 589.6 & 0.059 & \cellcolor{lightgray}\textbf{2.17} & 972.3 & 0.055 & 2.18 & 597.1 \\
GPT4o-mini & 0.055 & 1.08 & 1021.1 & 0.062 & 2.23 & 579.2 & 0.063 & 1.08 & 714.9 & 0.055 & 2.18 & 597.1 & 0.053 & 2.18 & 634.7 \\
GPT-4-Turbo & 0.053 & 2.97 & 993.5 & 0.067 & 2.28 & 737.6 & 0.065 & 1.22 & 810.3 & 0.055 & 2.18 & \cellcolor{lightgray}\textbf{273.5} & 0.029 & 2.25 & 366.8 \\
GPT-4o & 0.053 & 2.97 & 1002.5& 0.056& 2.25& 549.4& 0.052 & 1.35 & 589.6 & 0.055 & 2.18 & \cellcolor{lightgray}\textbf{273.5} & 0.055 & 2.18 & 597.1 \\
GPT-O1 & 0.044& 1.17& 910.7& 0.056& 2.25& 549.4& 0.052 & 1.35 & 589.6 & 0.055& 2.18& 597.1& 0.055 & 2.18 & 597.1 \\
RTLCoder-DS & 0.042 & 1.21 & 833.0 & 0.056 & 2.25 & 549.4 & 0.052 & 1.35 & 589.6 & 0.055 & 2.18 & 597.1 & 0.063& 2.34& 649.78\\
RTLrewriter & \cellcolor{lightgray}\textbf{0.025} & 3.23 & 424.0 & - & - & - & 0.041 & 1.36 & 549.4 & - & - & - & - & - & - \\
\name & 0.029 & \cellcolor{lightgray}\textbf{1.21} & \cellcolor{lightgray}\textbf{403.9}& \cellcolor{lightgray}\textbf{0.024}& \cellcolor{lightgray}\textbf{1.17} & \cellcolor{lightgray}\textbf{271.0}& \cellcolor{lightgray}\textbf{0.023} & \cellcolor{lightgray}\textbf{1.15}& \cellcolor{lightgray}\textbf{268.5} & \cellcolor{lightgray}\textbf{0.024} & \cellcolor{lightgray}\textbf{2.17} & \cellcolor{lightgray}\textbf{273.5} & \cellcolor{lightgray}\textbf{0.026} & 2.18 & \cellcolor{lightgray}\textbf{270.9}\\\hline
\textbf{ Improvement(\%)} & $\textcolor{green}{\boldsymbol{\Downarrow}}$30.95& 0.00& $\textcolor{green}{\boldsymbol{\Downarrow}}$51.51& $\textcolor{green}{\boldsymbol{\Downarrow}}$57.14& $\textcolor{green}{\boldsymbol{\Downarrow}}$48.00& $\textcolor{green}{\boldsymbol{\Downarrow}}$50.67& $\textcolor{green}{\boldsymbol{\Downarrow}}$55.77& $\textcolor{green}{\boldsymbol{\Downarrow}}$14.81& $\textcolor{green}{\boldsymbol{\Downarrow}}$54.46& $\textcolor{green}{\boldsymbol{\Downarrow}}$56.36& $\textcolor{green}{\boldsymbol{\Downarrow}}$0.46& $\textcolor{green}{\boldsymbol{\Downarrow}}$54.1952& $\textcolor{green}{\boldsymbol{\Downarrow}}$52.73& 0.00&$\textcolor{green}{\boldsymbol{\Downarrow}}$54.63\\ \Xhline{2.5\arrayrulewidth} 
 Original + Compiler Opt.& 0.041& 2.85& 564.51& 0.035& 2.24& 316.11& 0.038& 1.23& 358.77& 0.044& 2.19& 451.59& 0.045& 2.19&451.59\\
  \name + Compiler Opt.&\cellcolor{lightgray}\textbf{ 0.021}&\cellcolor{lightgray} \textbf{2.64}&\cellcolor{lightgray}\textbf{ 240.85}&\cellcolor{lightgray}\textbf{ 0.018}&\cellcolor{lightgray} \textbf{0.6}&\cellcolor{lightgray} \textbf{175.61}&\cellcolor{lightgray} \textbf{0.018}& 2.42&\cellcolor{lightgray} \textbf{180.63}&\cellcolor{lightgray} \textbf{0.020}&\cellcolor{lightgray} \textbf{2.17}&\cellcolor{lightgray} \textbf{185.65}&\cellcolor{lightgray} \textbf{0.019}& 2.27&\cellcolor{lightgray}\textbf{188.169}\\
\hline
\textbf{Improvement(\%)} & $\textcolor{green}{\boldsymbol{\Downarrow}}$48.78& $\textcolor{green}{\boldsymbol{\Downarrow}}$7.37& $\textcolor{green}{\boldsymbol{\Downarrow}}$57.33& $\textcolor{green}{\boldsymbol{\Downarrow}}$48.57& $\textcolor{green}{\boldsymbol{\Downarrow}}$73.21& $\textcolor{green}{\boldsymbol{\Downarrow}}$44.45& $\textcolor{green}{\boldsymbol{\Downarrow}}$52.63& $\textcolor{red}{\boldsymbol{\Uparrow}}$96.74& $\textcolor{green}{\boldsymbol{\Downarrow}}$49.65& $\textcolor{green}{\boldsymbol{\Downarrow}}$54.55& $\textcolor{green}{\boldsymbol{\Downarrow}}$0.91& $\textcolor{green}{\boldsymbol{\Downarrow}}$58.89& $\textcolor{green}{\boldsymbol{\Downarrow}}$57.78& $\textcolor{red}{\boldsymbol{\Uparrow}}$3.65& $\textcolor{green}{\boldsymbol{\Downarrow}}$58.33\\
\Xhline{2.5\arrayrulewidth} 
\end{tabular}
}
\end{table*}

\setcounter{table}{6} 
\begin{table*}[t]
\caption{Algorithm Optimizations PPA Comparison. Gray highlights indicate state-of-the-art results. A $\textcolor{green}{\boldsymbol{\Downarrow}}$ marks improvement, while a $\textcolor{red}{\boldsymbol{\Uparrow}}$ denotes a decline compare with the original design. Two comparison scenarios are shown: without compiler optimization (upper improvement) and with compiler optimization (lower improvement).}
\label{tab:algo-comparison}
\setlength{\tabcolsep}{1pt} 
\centering
\resizebox{1\textwidth}{!}{
\fontsize{8}{7}\selectfont
\renewcommand{\arraystretch}{1.3} 
\begin{tabular}{l|ccc|ccc|ccc|ccc|ccc}
\Xhline{2.5\arrayrulewidth} 
\textbf{Model/Method} & \multicolumn{3}{c|}{\textbf{sppm\_redundancy}} & \multicolumn{3}{c|}{\textbf{subexpression\_elim}} & \multicolumn{3}{c|}{\textbf{adder\_architecture}} &  \multicolumn{3}{c|}{\textbf{vending}}&\multicolumn{3}{c}{\textbf{fft}} \\
& \begin{tabular}[c]{@{}c@{}}Power\\(mW)\end{tabular} & \begin{tabular}[c]{@{}c@{}}Time\\(ns)\end{tabular} & \begin{tabular}[c]{@{}c@{}}Area\\($\mu\text{m}^2$)\end{tabular} & 
\begin{tabular}[c]{@{}c@{}}Power\\(mW)\end{tabular} & \begin{tabular}[c]{@{}c@{}}Time\\(ns)\end{tabular} & \begin{tabular}[c]{@{}c@{}}Area\\($\mu\text{m}^2$)\end{tabular} & 
\begin{tabular}[c]{@{}c@{}}Power\\(mW)\end{tabular} & \begin{tabular}[c]{@{}c@{}}Time\\(ns)\end{tabular} & \begin{tabular}[c]{@{}c@{}}Area\\($\mu\text{m}^2$)\end{tabular} & 
 \begin{tabular}[c]{@{}c@{}}Power\\(mW)\end{tabular} & \begin{tabular}[c]{@{}c@{}}Time\\(ns)\end{tabular} & \begin{tabular}[c]{@{}c@{}}Area\\($\mu\text{m}^2$)\end{tabular} &\begin{tabular}[c]{@{}c@{}}Power\\(mW)\end{tabular} & \begin{tabular}[c]{@{}c@{}}Time\\(ns)\end{tabular} & \begin{tabular}[c]{@{}c@{}}Area\\($\mu\text{m}^2$)\end{tabular} \\
\hline
Original & 2.86& 7.41 & 40102.6 & 5.27& 11.09 & 10989.1 & 0.418 & 2.78 & 1023.5 &  7.61& 227.86& 176982.98&58.23& 8.26& 2255264.75\\
GPT-3.5-Turbo & 2.86& 7.41 & 40102.6 & 5.27& 11.09 & 10989.1 & 0.418 & 2.78 & 1023.5 &  7.61& 227.86& 176982.98&58.23& 8.26& 2255264.75\\
GPT4o-mini & 2.86& 7.41 & 40102.6 & 5.27& 11.09 & 10989.1 & 0.418 & 2.78 & 1023.5 &  7.61& 227.86& 176982.98&58.23& 8.26& 2255264.75\\
GPT-4-Turbo & 2.86& 7.41& 40102.6& 3.93& 11.09 & 7783.03 & 0.392 & 2.74 & 1023.5 &  7.50& 227.86& 176982.98&58.23& 8.26& 2255264.75\\
GPT-4o& 2.86& 7.41& 40102.6& 5.27& 11.09 & 8984.61 & 0.392& 2.74 & 1023.5 & 7.61& 227.86& 176982.98& 58.23& 8.26&2255264.75\\
GPT-O1& 1.87& 7.41& 29919.8& 4.63& 11.09& 8957.23& 0.418& 2.78& 1023.52& 7.61& 227.86& 176982.98& 58.23& 8.26&2255264.75\\
RTLCoder-DS & 2.86& 7.41 & 40102.6 & 5.27& 11.09 & 10989.1 & 0.418 & 2.78 & 1023.5 &  7.61& 227.86& 176982.98&58.23& 8.26& 2255264.75\\
\name & \cellcolor{lightgray}\textbf{1.77}& \cellcolor{lightgray}\textbf{7.29} & \cellcolor{lightgray}\textbf{29606.18} & \cellcolor{lightgray}\textbf{3.02}& \cellcolor{lightgray}\textbf{2.87}& \cellcolor{lightgray}\textbf{7358.8}& \cellcolor{lightgray}\textbf{0.328} & \cellcolor{lightgray}\textbf{1.97} & \cellcolor{lightgray}\textbf{762.6} &  \cellcolor{lightgray}\textbf{6.97}& 227.86& \cellcolor{lightgray}\textbf{164831.1}&\cellcolor{lightgray}\textbf{31.71}& \cellcolor{lightgray}\textbf{8.09}& \cellcolor{lightgray}\textbf{1726125.71}\\\hline 
\textbf{ Improvement(\%)}& $\textcolor{green}{\boldsymbol{\Downarrow}}$38.46& $\textcolor{green}{\boldsymbol{\Downarrow}}$1.62& $\textcolor{green}{\boldsymbol{\Downarrow}}$26.17& $\textcolor{green}{\boldsymbol{\Downarrow}}$42.68& $\textcolor{green}{\boldsymbol{\Downarrow}}$74.12& $\textcolor{green}{\boldsymbol{\Downarrow}}$33.04& $\textcolor{green}{\boldsymbol{\Downarrow}}$21.53& $\textcolor{green}{\boldsymbol{\Downarrow}}$29.14& $\textcolor{green}{\boldsymbol{\Downarrow}}$25.49& $\textcolor{green}{\boldsymbol{\Downarrow}}$8.41& $\textcolor{green}{\boldsymbol{\Downarrow}}$0& $\textcolor{green}{\boldsymbol{\Downarrow}}$6.87& $\textcolor{green}{\boldsymbol{\Downarrow}}$45.54& $\textcolor{green}{\boldsymbol{\Downarrow}}$2.06&$\textcolor{green}{\boldsymbol{\Downarrow}}$23.46\\\Xhline{2.5\arrayrulewidth}
 Original + Compiler Opt.& 1.46& 7.95& 22908.69& 4.61& 11.78& 9484.15& 0.17& 2.29& 541.92& 11.46& 7.90& 240079.29& 51.12& 7.90&1857805.49\\
  \name + Compiler Opt.& 1.46& 7.95& 22908.69& \cellcolor{lightgray}\textbf{3.53}& 11.78& \cellcolor{lightgray} \textbf{6791.88}& 0.17& 2.48& \cellcolor{lightgray} \textbf{531.88}& \cellcolor{lightgray} \textbf{8.175}& 7.90& \cellcolor{lightgray} \textbf{151593.86} & \cellcolor{lightgray} \textbf{26.32}& 8.98& \cellcolor{lightgray}\textbf{1471378.46}\\
\hline
\textbf{Improvement(\%)}& 0& 0& 0& $\textcolor{green}{\boldsymbol{\Downarrow}}$23.4 & 0 & $\textcolor{green}{\boldsymbol{\Downarrow}}$28.39& 0 & $\textcolor{red}{\boldsymbol{\Uparrow}}$8.30 & $\textcolor{green}{\boldsymbol{\Downarrow}}$1.86&  $\textcolor{green}{\boldsymbol{\Downarrow}}$28.66 &0 & $\textcolor{green}{\boldsymbol{\Downarrow}}$36.86 &$\textcolor{green}{\boldsymbol{\Downarrow}}$48.51 & $\textcolor{red}{\boldsymbol{\Uparrow}}$13.67& $\textcolor{green}{\boldsymbol{\Downarrow}}$20.8 \\
\Xhline{2.5\arrayrulewidth} 
\end{tabular}
}
\end{table*}

\begin{wrapfigure}{r}{0.6\textwidth}
     \centering    
     \includegraphics[width=0.6\textwidth]{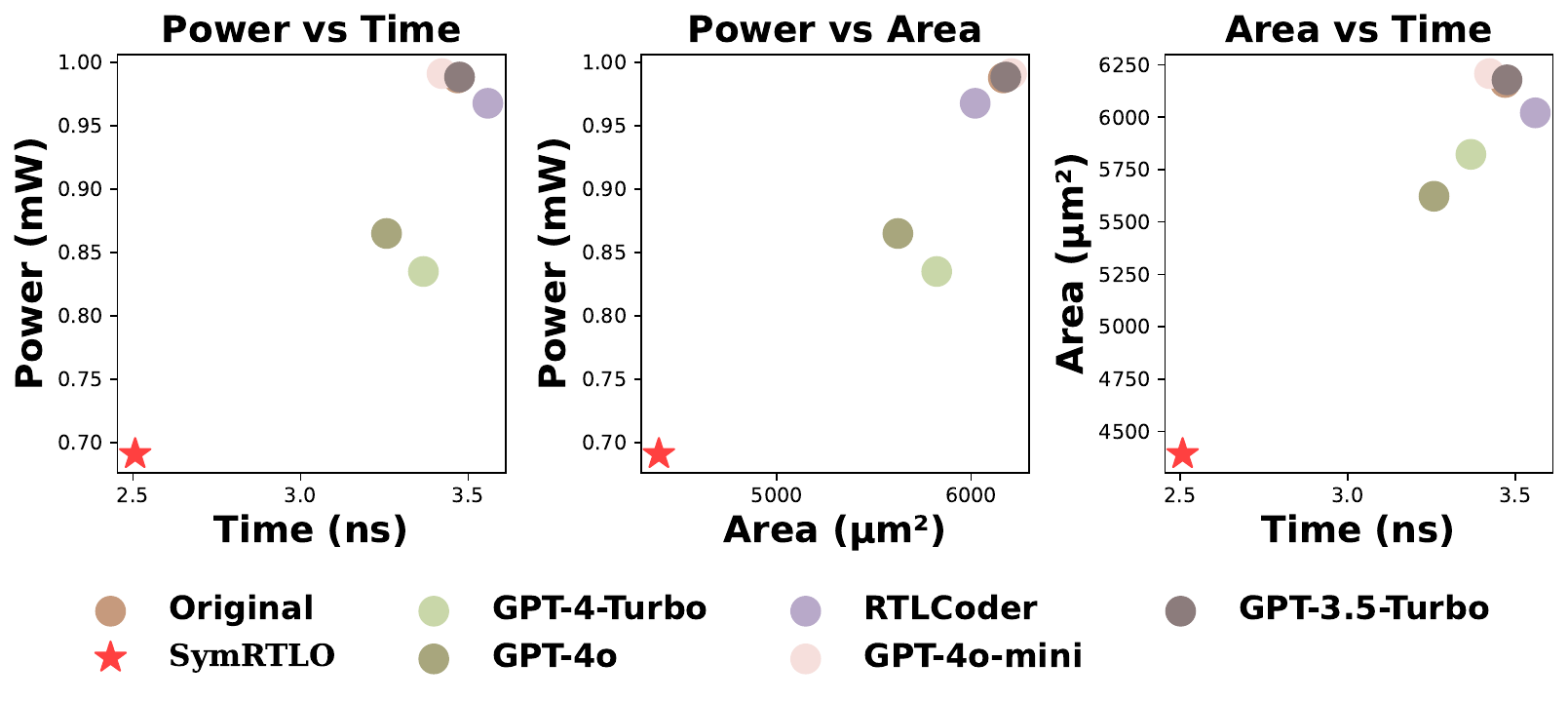}
     \caption{The PPA overall improvement of benchmark cases.}
     \label{fig:comparison}
\end{wrapfigure}

% optimization goal reached - improve area as primary goal? area: because the examples in the benchmark lack most of critera
To demonstrate \name's effectiveness in resolving cross-rule conflicts and achieving optimization objectives, we conduct an experiment presented in Figure \ref{fig:comparison}. Given the limited benchmarks available for LLM-driven RTL design, we analyze our framework on RTLRewriter's benchmark, which primarily emphasizes area optimization, leaving minimal room for improvements in power and delay. To align with this limitation, we put area optimization as our primary goal. Despite these constraints, \name achieves substantial improvements, averaging 40.96\% in power, 17.02\% in delay, and 38.05\% in area, while maintaining a balanced optimization across all three metrics, highlighting its versatility and robustness.

We present a comprehensive analysis of  \name capabilities using 11 short-benchmark examples from RTLRewriter focusing on Wires and Cells optimization, along with 10 complex FSM and algorithm examples from both short and long benchmarks. This representative selection demonstrates \name's scalability and effectiveness across diverse functional domains but also aligns with circuits reported in RTLRewriter's paper, enabling direct comparison with state-of-the-art results.

% optimization alignment with optimizaiton algorithm
Smaller benchmarks requiring only 1-2 optimization patterns provide ideal test cases for LLM output \textit{alignment}.  Table \ref{tab:model-comparison} shows that \name consistently outperforms baseline implementations across various test cases. With wire and cell ratios of 0.63 and 0.67 respectively, it surpasses the state-of-the-art values of 0.69 and 0.77. While models like GPT-4 excel in certain cases, they lack consistency across diverse optimization tasks.

%Even RTL-focused models with specialized training exhibit limitations in optimization tasks.

In FSM PPA experiments, \name significantly outperforms existing approaches, particularly in relation to RTLRewriter, the state-of-the-art solution, achieving an improvement of up to 50.59\%, 12.65\%, 53.09\% in power, time, and area, respectively. As shown in Table \ref{tab:fsm-comparison}, it effectively aligns the FSM state reduction algorithm with optimized code, minimizing all FSM states and achieving the best overall PPA results. This demonstrates that the LLM-generated symbolic system is both stable and aligned with intended optimization goals.
%optimization alignment, optimization generalization( optimization goal prompt ), goal target conflicts, AST template usage

To evaluate the effectiveness of generalized rules and AST templates in balancing conflicting rules, we conduct algorithm case PPA experiments involving complex Data Path and Control Path scenarios. As shown in Table \ref{tab:algo-comparison}, \name applies AST templates, optimized rules, and minimized FSM states, achieving 30.34\%, 21.37\%, and 20.01\% improvements in PPA over GPT-4o,  our base model, on average. Note that RTLRewriter's results are unavailable for comparison for these cases.

% We have run the PPA results with Synopsys DC optimization workflows in both FSM and Algorithm test cases, as shown in Table \ref{tab:fsm-comparison} and Table \ref{tab:algo-comparison}. We have shown that \name can achieve a balanced optimization in the scenario with compiler optimization, which makes \name an essential tool in the EDA toolchain. 

% In both short- and long-reference cases, our framework consistently outperformed the state-of-the-art, particularly excelling in complex scenarios. The comparison of \name's performance within the standard compiler optimization workflow is impressive, demonstrating significant improvements in PPA results compared to the baseline. This underscores the benefits of integrating \name into the standard optimization automation process. Moreover, \name demonstrated remarkable stability, producing reliable and consistent results in a variety of circuits and input sizes, from simple combinational logic to complex state machines. This contrasts with other approaches, including traditional compiler-based methods and LLMs, which often show variable performance across different optimization scenarios. A key distinguishing feature of \name is its goal-based approach to optimizing PPA.  This approach enables simultaneous improvements without introducing pattern conflicts, causing confusion for LLM. 

We test \name with Synopsys DC optimization workflows with medium mapping effort, incremental mapping for both FSM and Algorithm cases, as shown in Table \ref{tab:fsm-comparison} and Table \ref{tab:algo-comparison}, demonstrating further balanced optimization alongside compiler optimization processes, achieving overall improvements of 36.2\% in power and 35.66\% in area, with only an 8.3\% increase in time as a trade-off. Even under more stringent compiler optimization settings of flattened mode with high mapping effort, \name still delivers overall improvements of 27.7\% in area, 35.8\% in power, and 0.5\% in delay.

% Across short- and long-reference cases, our framework consistently outperformed state-of-the-art methods, especially in complex scenarios. \name showed significant PPA improvements within standard compiler optimization workflows, highlighting the benefits of integrating it into the optimization automation process. Additionally, \name exhibited remarkable stability, delivering reliable results across various circuits and input sizes—from simple combinational logic to complex state machines—unlike traditional compiler methods and LLMs, which often show variable performance. A key feature of \name is its goal-based PPA optimization approach, enabling simultaneous improvements without pattern conflicts that typically confuse LLMs.

% \begin{table}[htbp]
% \centering

% \begin{minipage}{0.45\textwidth}
% \centering
% {\fontsize{7}{9}\selectfont
% \begin{tabular}{l|c|c|c|c}
% \hline
% \textbf{Component} & \textbf{Area} & \textbf{Power} & \textbf{Timing} & \textbf{Functional} \\
% & \textbf{Reduction (\%)} & \textbf{Reduction (\%)} & \textbf{Improvement (\%)} & \textbf{Errors (\%)} \\ \hline
% Full Framework & \textbf{47} & \textbf{48} & \textbf{91} & \textbf{0} \\
% Without Templates & 26 & 22 & 54 & 12 \\
% Without Symbolic Reasoning & 38 & 35 & 72 & 18 \\
% Without Goal-Based Search & 31 & 43 & 60 & 7 \\ \hline
% \end{tabular}
% }
% \caption{Impact of ablation on area, power, timing, and functional errors.}
% \label{tab:ablation}
% \end{minipage}
% \end{table}

\vspace{-5pt}
\vspace{-5pt}
\subsection{Ablation Studies}
\vspace{-5pt}

\begin{wrapfigure}{r}{0.6\textwidth}
    \centering
    \includegraphics[width=1\linewidth]{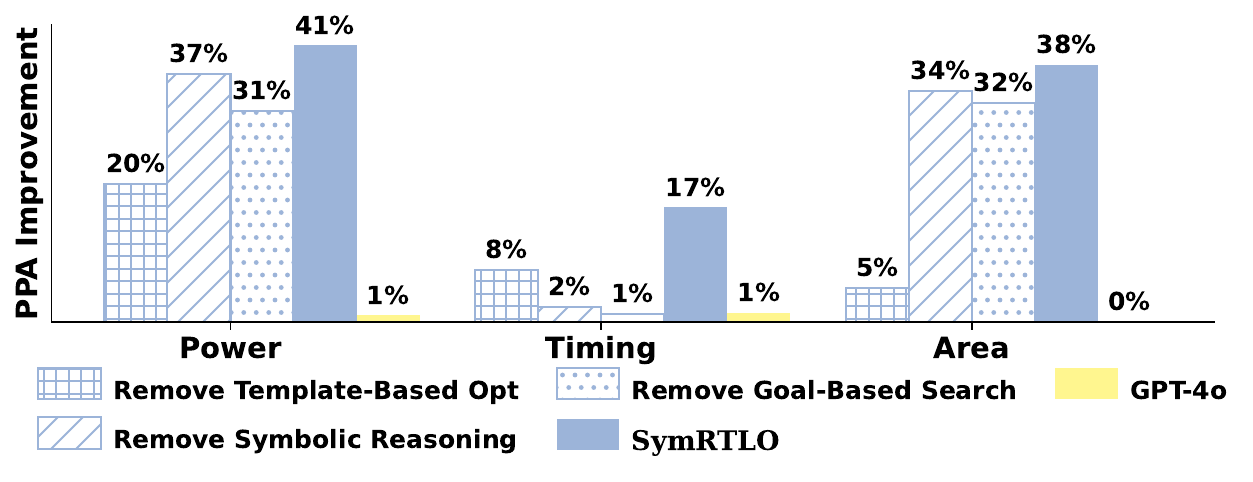}
    
    \caption{Ablation studies results.}
    \label{fig:ablation}
\end{wrapfigure}

To assess the effectiveness of individual components in \name, we conduct ablation studies by selectively removing one module at a time, \ie either the AST-based module, FSM symbolic system, or goal-based search engine, while keeping the rest of the system intact. We then analyze the impact of each removal across test cases, measuring the resulting overall PPA improvements compared to the baseline. Figure~\ref{fig:ablation}  summarizes these results, showing that all three components contribute significantly to \name's overall performance. Removing any one of them results in significant performance degradation, further emphasizing the necessity of their integration. By contrast, GPT-4o alone achieves minimal improvements, underscoring the advantages of \name's tailored framework.

%% file: sections/conclusion.tex
\vspace{-10pt}
\section{Conclusion}
\vspace{-7pt}
We present \name, a neuron-symbolic framework that integrates LLM-based code rewriting and symbolic reasoning to optimize both data flow and control flow in RTL designs. \name generalizes optimization rules, aligns generated code with intended transformations, resolves conflicting optimization goals, and ensures reliable automated verification. By combining retrieval-augmented guidance with symbolic systems, \name automates complex structural rewrites while maintaining functional correctness. Extensive evaluations on industrial-scale designs demonstrate significant PPA gains over state-of-the-art solutions.

%% file: sections/appendix.tex
\onecolumn
% Required packages
\renewcommand{\normalsize}{\fontsize{10}{12}\selectfont}

% Define code box with fixed syntax
% Define code box with fixed syntax
% Define code box with fixed syntax
\newtcblisting[auto counter]{unifiedcode}[4][]{%
enhanced jigsaw,
interior style={white},
frame style={black},
colback=white,
colframe=black,
arc=0pt,
boxrule=#3,
toprule=#3,
bottomrule=#3,
leftrule=#3,
rightrule=#3,
left=2mm,
right=2mm,
top=2mm,
bottom=1mm,
boxsep=1pt,
title style={color=white},
attach boxed title to top left={yshift=-0.5mm},
boxed title style={
    colback=black,
    size=small,
    colframe=black,
    boxrule=0pt
},
listing engine=minted,
minted language=#1,
minted style=default,
minted options={
    linenos,
    numbersep=1pt,
    breaklines=true,
    autogobble,
    fontsize=\footnotesize,
    baselinestretch=0.9,
    frame=none
},
after={\vspace{0.5mm}\begin{center}\textit{Listing \thetcbcounter:} #2\end{center}\vspace{0.5mm}},
size=title,
listing only
}

\section{FSM Symbolic System}\label{app:fsm}
\vspace{-15pt}

The following section demonstrates an example FSM in verilog. First the verilog is transformed to Symbolic Representation, then the Symbolic system applied minimization algorithm to optimize the FSM.

\vspace{-5pt}

\begin{unifiedcode}[verilog]{Example Test Case: example1\_state.}{0pt}{gray!50}
module example(
    input wire clk,
    input wire reset,
    input wire [1:0] input_signal, 
    output reg output_signal);
    parameter S0 = 3'b000,S1 = 3'b001,S2 = 3'b010,S3 = 3'b011,S4 = 3'b100,S5 = 3'b101;
    reg [2:0] current_state, next_state;
    always @(current_state) begin
        output_signal = 0;
        case (current_state)
            S0: output_signal = 1;
            S2: output_signal = 1;
            S4: output_signal = 1;
            default: output_signal = 0;
        endcase
    end
    always @(posedge clk or posedge reset) begin
        if (reset) begin
            current_state <= S0; // Reset to state S0
        end else begin
            current_state <= next_state;
        end
    end
    always @(*) begin
        next_state = current_state;
        case (current_state)
            S0: case (input_signal)
                2'b00: next_state = S0;
                2'b01: next_state = S1;
                2'b10: next_state = S2;
                2'b11: next_state = S3;
            endcase
            S1: case (input_signal)
                2'b00: next_state = S0;
                2'b01: next_state = S3;
                2'b11: next_state = S5;
            endcase
            S2: case (input_signal)
                2'b00: next_state = S1;
                2'b01: next_state = S3;
                2'b10: next_state = S2;
                2'b11: next_state = S4;
            endcase
            S3: case (input_signal)
                2'b00: next_state = S1;
                2'b01: next_state = S0;
                2'b10: next_state = S4;
                2'b11: next_state = S5;
            endcase
            S4: case (input_signal)
                2'b00: next_state = S0;
                2'b01: next_state = S1;
                2'b10: next_state = S2;
                2'b11: next_state = S5;
            endcase
            S5: case (input_signal)
                2'b00: next_state = S1;
                2'b01: next_state = S4;
                2'b10: next_state = S0;
            endcase
        endcase
    end
endmodule
\end{unifiedcode}

% For JSON
\begin{unifiedcode}[Python]{LLM-generated symbolic representation for example1\_state.}{1pt}{gray!50}
    'states': ['S0', 'S1', 'S2', 'S3', 'S4', 'S5'],
    'transitions': {
        'S0': {
            'input_signal=00': {'next_state': 'S0'},
            'input_signal=01': {'next_state': 'S1'},
            'input_signal=10': {'next_state': 'S2'},
            'input_signal=11': {'next_state': 'S3'}
        },
        'S1': {
            'input_signal=00': {'next_state': 'S0'},
            'input_signal=01': {'next_state': 'S3'},
            'input_signal=11': {'next_state': 'S5'}
        },
        'S2': {
            'input_signal=00': {'next_state': 'S1'},
            'input_signal=01': {'next_state': 'S3'},
            'input_signal=10': {'next_state': 'S2'},
            'input_signal=11': {'next_state': 'S4'}
        },
        'S3': {
            'input_signal=00': {'next_state': 'S1'},
            'input_signal=01': {'next_state': 'S0'},
            'input_signal=10': {'next_state': 'S4'},
            'input_signal=11': {'next_state': 'S5'}
        },
        'S4': {
            'input_signal=00': {'next_state': 'S0'},
            'input_signal=01': {'next_state': 'S1'},
            'input_signal=10': {'next_state': 'S2'},
            'input_signal=11': {'next_state': 'S5'}
        },
        'S5': {
            'input_signal=00': {'next_state': 'S1'},
            'input_signal=01': {'next_state': 'S4'},
            'input_signal=10': {'next_state': 'S0'}
        }
    },
    'outputs': {'S0': {'output_signal': 1},
                'S1': {'output_signal': 0},
                'S2': {'output_signal': 1},
                'S3': {'output_signal': 0},
                'S4': {'output_signal': 1},
                'S5': {'output_signal': 0} }
\end{unifiedcode}

% For final states
\begin{unifiedcode}[Python]{Reduced states of example1\_state.}{0.5pt}{gray!30}
State: S2, Output: 1
  input_signal=00 -> S1
  input_signal=01 -> S3_S5
  input_signal=10 -> S2
  input_signal=11 -> S0_S4
State: S0_S4, Output: 1
  input_signal=00 -> S0_S4
  input_signal=01 -> S1
  input_signal=10 -> S2
  input_signal=11 -> S3_S5
State: S1, Output: 0
  input_signal=00 -> S0_S4
  input_signal=01 -> S3_S5
  input_signal=10 -> S1
  input_signal=11 -> S3_S5
State: S3_S5, Output: 0
  input_signal=00 -> S1
  input_signal=01 -> S0_S4
  input_signal=10 -> S0_S4
  input_signal=11 -> S3_S5
\end{unifiedcode}

\section{AST Template}\label{app:ast}

The following section demonstrate how applying AST templates transforms the code and apply optimization patterns. 
\begin{unifiedcode}[verilog]{Example Test Case: dead\_code\_elimination.}{1pt}{black}
module example_raw
#(  parameter       BW = 8)
(
    input [BW-1:0] a,
    input [BW-1:0] b,
    input [BW-1:0] c,
    input [BW-1:0] d,
    output [BW-1:0] s1
);
    assign s2 = a * b;
    assign s3 = a % b +d;
    assign s4 = c + d + b * a;
    assign s5 = a - b;
    assign s6 = (b + 1) * a + d + c -b;
    assign s1 = a + 23;
endmodulee
\end{unifiedcode}

\vspace{-5pt}

\begin{unifiedcode}[verilog]{Example Test Case: dead\_code\_elimination after applying the Dead Code Elimination AST template.}{1pt}{black}
module example_raw #
(parameter BW = 8)
(
  input [BW-1:0] a,
  input [BW-1:0] b,
  input [BW-1:0] c,
  input [BW-1:0] d,
  output [BW-1:0] s1 
);
  assign s1 = a + 23;
endmodule

\end{unifiedcode}
\vspace{-8pt}
\begin{unifiedcode}[verilog]{Example Test Case: subexpression\_elimination.}{1pt}{black}
module example_raw
#(  parameter       BW = 8)
(
    input [BW-1:0] a,
    input [BW-1:0] b,
    input [BW-1:0] c,
    input [BW-1:0] d,
    output [BW-1:0] s1,
    output [BW-1:0] s2,
    output [BW-1:0] s3,
    output [BW-1:0] s4,
    output [BW-1:0] s5,
    output [BW-1:0] s6 
);
    assign s1 = a + b;
    assign s2 = a * b;
    assign s3 = a \% b +d;
    assign s4 = c + d + b * a;
    assign s5 = a - b;
    assign s6 = (b + 1) * a + d + c -b;
endmodule
\end{unifiedcode}
\vspace{-5pt}
\begin{unifiedcode}[verilog]{Example Test Case: subexpression\_elimination after applying the Common Sub-Expressions Elimination template. The Common Sub-Expressions are reused in the states after it. }{1pt}{black}
module example
#(  parameter       BW = 8)
(   input [BW-1:0] a,
    input [BW-1:0] b,
    input [BW-1:0] c,
    input [BW-1:0] d,
    output [BW-1:0] s1,
    output [BW-1:0] s2,
    output [BW-1:0] s3,
    output [BW-1:0] s4,
    output [BW-1:0] s5,
    output [BW-1:0] s6 
);
    assign s1 = a + b;
    assign s2 = a * b;
    assign s3 = a \% b + d;
    assign s4 = c + d + s2;
    assign s5 = a - b;
    assign s6 = s4 + s5;
endmodule
\end{unifiedcode}

\vspace{-5pt}
\begin{unifiedcode}[verilog]{Example Test Case: algebraic\_simplification.}{1pt}{black}
module example_raw
#(  parameter    BW = 8)
(
    input [BW-1:0] a,
    input [BW-1:0] b,
    output [BW-1:0] s1,
    output [BW-1:0] s2
);
    wire [BW-1:0] t1, t2;
    assign s1 = a + b;
    assign t1 = s1 + 0;
    assign t2 = s1 * 1;
    assign s2 = t1 + t2;
endmodule

\end{unifiedcode}
\vspace{-5pt}
\begin{unifiedcode}[verilog]{Example Test Case: algebraic\_simplification after applying the Temporary Variable Elimination, Dead Code Elimination, then Expression Simplification templates.}{1pt}{black}

module example_raw #
(
  parameter BW = 8
)
(
  input [BW-1:0] a,
  input [BW-1:0] b,
  output [BW-1:0] s1,
  output [BW-1:0] s2
);

  assign s1 = a + b;
  assign s2 = s1 + s1;

endmodule
\end{unifiedcode}

\section{RAG Example}\label{app:rag}
\begin{figure}[H]
    \centering
    \begin{mdframed}[
        linewidth=0.5pt,
        roundcorner=4pt,
        backgroundcolor=blue!3,
    ]
        \textcolor{black}{\textbf{Sample Retrieval Augmented Optimization Rule}} \\[0.5em]
        \textcolor{blue}{\texttt{"}\textrm{name}\texttt{":}} \textcolor{teal}{\texttt{"}\textrm{Zero Multiplication Elimination}\texttt{"}},

        \textcolor{blue}{\texttt{"}\textrm{pattern}\texttt{":}} \textcolor{teal}{\texttt{"}\textrm{Detect multiplication by zero in expressions (\eg, 0 * c)}\texttt{"}},
        
        \textcolor{blue}{\texttt{"}\textrm{rewrite}\texttt{":}} \textcolor{teal}{\texttt{"}\textrm{Eliminate multiplication by zero, replacing the entire expression with zero}\texttt{"}},
        
        \textcolor{blue}{\texttt{"}\textrm{category}\texttt{":}} \textcolor{teal}{\texttt{"}\textrm{combinational/dataflow}\texttt{"}},
        
        \textcolor{blue}{\texttt{"}\textrm{objective\_improvement}\texttt{":}} \textcolor{teal}{\texttt{"}\textrm{area}\texttt{"}},
        
        \textcolor{blue}{\texttt{"}\textrm{template\_guidance}\texttt{":}} \textcolor{teal}{\texttt{"}\textrm{Identify vast.Times nodes with a zero operand. Replace the node with a vast.IntConst node representing zero.}\texttt{"}},
        
        \textcolor{blue}{\texttt{"}\textrm{function\_name}\texttt{":}} \textcolor{teal}{\texttt{"}\textrm{ZeroMultiplicationTemplate}\texttt{"}}
    \end{mdframed}
    \caption{Sample Retrieval Augmented Optimization Rule 1: Zero Multiplication Rule.}
    \label{fig:system-prompt}
\end{figure}

\begin{figure}[H]
\centering
\begin{mdframed}[
linewidth=0.5pt,
roundcorner=4pt,
backgroundcolor=blue!3,
]
\textcolor{black}{\textbf{Sample Retrieval Augmented Optimization Rule}} \\[0.5em]
\textcolor{blue}{\texttt{"}\textrm{name}\texttt{":}} \textcolor{teal}{\texttt{"}\textrm{IntermediateVariableExtraction}\texttt{"}},
    
    \textcolor{blue}{\texttt{"}\textrm{pattern}\texttt{":}} \textcolor{teal}{\texttt{"}\textrm{Detect conditional assignments to a register based on a control signal}\texttt{"}},
    
    \textcolor{blue}{\texttt{"}\textrm{rewrite}\texttt{":}} \textcolor{teal}{\texttt{"}\textrm{Extract common sub-expressions into intermediate variables to reduce redundant logic}\texttt{"}},
    
    \textcolor{blue}{\texttt{"}\textrm{category}\texttt{":}} \textcolor{teal}{\texttt{"}\textrm{combinational/dataflow}\texttt{"}},
    
    \textcolor{blue}{\texttt{"}\textrm{objective\_improvement}\texttt{":}} \textcolor{teal}{\texttt{"}\textrm{area}\texttt{"}},
    
    \textcolor{blue}{\texttt{"}\textrm{template\_guidance}\texttt{":}} \textcolor{teal}{\texttt{"}\textrm{To implement this rule in a Python template subclassing BaseTemplate, use pyverilog AST manipulation to identify conditional assignments (vast.IfStatement) and extract the common sub-expressions into separate assignments. Look for vast.Identifier nodes that are assigned conditionally and create new vast.Assign nodes for the intermediate variables. Ensure that the new assignments are placed before the conditional logic to maintain correct data flow}\texttt{"}},
    
    \textcolor{blue}{\texttt{"}\textrm{function\_name}\texttt{":}} \textcolor{teal}{\texttt{"}\textrm{IntermediateVariableExtractionTemplate}\texttt{"}}
\end{mdframed}
\caption{Sample Retrieval Augmented Optimization Rule 2: Intermediate Variable Extraction.}
\label{fig:intermediate-variable}
\end{figure}

\begin{figure}[H]
\centering
\begin{mdframed}[
linewidth=0.5pt,
roundcorner=4pt,
backgroundcolor=blue!3,
]
\textcolor{black}{\textbf{Hardware Optimization Rule}} \\[0.5em]
\textcolor{blue}{\texttt{"}\textrm{name}\texttt{":}} \textcolor{teal}{\texttt{"}\textrm{ReplaceRippleCarryWithCarryLookahead}\texttt{"}},
    
    \textcolor{blue}{\texttt{"}\textrm{pattern}\texttt{":}} \textcolor{teal}{\texttt{"}\textrm{Detects a ripple carry adder implementation using a series of full adders connected in sequence}\texttt{"}},
    
    \textcolor{blue}{\texttt{"}\textrm{rewrite}\texttt{":}} \textcolor{teal}{\texttt{"}\textrm{Transforms the ripple carry adder into a carry lookahead adder by using partial full adders and generating carry bits in parallel}\texttt{"}},
    
    \textcolor{blue}{\texttt{"}\textrm{category}\texttt{":}} \textcolor{teal}{\texttt{"}\textrm{combinational/dataflow}\texttt{"}},
    
    \textcolor{blue}{\texttt{"}\textrm{objective\_improvement}\texttt{":}} \textcolor{teal}{\texttt{"}\textrm{area, delay}\texttt{"}},
    
    \textcolor{blue}{\texttt{"}\textrm{template\_guidance}\texttt{":}} \textcolor{teal}{\texttt{null}},
    
    \textcolor{blue}{\texttt{"}\textrm{function\_name}\texttt{":}} \textcolor{teal}{\texttt{null}}
\end{mdframed}
\caption{Sample Retrieval Augmented Optimization Rule 3: Replace Ripple Carry with Carry Lookahead, no template guidance is needed since it is an abstract rule.}
\label{fig:ripple-carry}
\end{figure}

%% file: main.bbl
\begin{thebibliography}{10}

\bibitem{blocklove:2023:chipchat}
Jason Blocklove, Siddharth Garg, Ramesh Karri, and Hammond Pearce.
\newblock Chip-chat: Challenges and opportunities in conversational hardware design.
\newblock In {\em 5th {ACM/IEEE} Workshop on Machine Learning for CAD, {MLCAD}}. {IEEE}, 2023.

\bibitem{BraytonMishchenko2010abc}
R.~Brayton and A.~Mishchenko.
\newblock {ABC}: An academic industrial-strength verification tool.
\newblock In {\em Computer Aided Verification: 22nd International Conference, {CAV} 2010, Edinburgh, UK, July 15--19, 2010. Proceedings 22}, pages 24--40. Springer, 2010.

\bibitem{buchberger1982algebraic}
Bruno Buchberger and R{\"u}diger Loos.
\newblock Algebraic simplification.
\newblock In {\em Computer algebra: symbolic and algebraic computation}, pages 11--43. Springer, 1982.

\bibitem{carette2004understanding}
Jacques Carette.
\newblock Understanding expression simplification.
\newblock In {\em Proceedings of the 2004 international symposium on Symbolic and algebraic computation}, pages 72--79, 2004.

\bibitem{chang:2023:chipgpt}
Kaiyan Chang, Ying Wang, Haimeng Ren, Mengdi Wang, Shengwen Liang, Yinhe Han, Huawei Li, and Xiaowei Li.
\newblock Chipgpt: How far are we from natural language hardware design.
\newblock {\em CoRR}, abs/2305.14019, 2023.

\bibitem{chen2004register}
Deming Chen and Jason Cong.
\newblock Register binding and port assignment for multiplexer optimization.
\newblock In {\em Proceedings of the 2004 Asia and South Pacific Design Automation Conference}, ASP-DAC '04, page 68–73. IEEE Press, 2004.

\bibitem{chen2024pyod2}
Sihan Chen, Zhuangzhuang Qian, Wingchun Siu, Xingcan Hu, Jiaqi Li, Shawn Li, Yuehan Qin, Tiankai Yang, Zhuo Xiao, Wanghao Ye, Yichi Zhang, Yushun Dong, and Yue Zhao.
\newblock Pyod 2: A python library for outlier detection with llm-powered model selection.
\newblock {\em arXiv}, 2412.12154, December 2024.

\bibitem{chu2006rtl}
Pong~P Chu.
\newblock {\em RTL hardware design using VHDL: coding for efficiency, portability, and scalability}.
\newblock John Wiley \& Sons, 2006.

\bibitem{cocke1970global}
John Cocke.
\newblock Global common subexpression elimination.
\newblock In {\em Proceedings of a symposium on Compiler Optimization}, pages 20--24. ACM, 1970.

\bibitem{cooper2001operator}
Keith~D Cooper, L~Taylor Simpson, and Christopher~A Vick.
\newblock Operator strength reduction.
\newblock {\em ACM Transactions on Programming Languages and Systems (TOPLAS)}, 23(5):603--625, 2001.

\bibitem{tilwani2024neurosymbolic}
Revathy~Venkataramanan Deepa~Tilwani and Amit~P. Sheth.
\newblock Neurosymbolic ai approach to attribution in large language models.
\newblock {\em arXiv}, 2410.03726, September 2024.
\newblock Paper under review.

\bibitem{calanzone2024neurosymbolic}
Stefano~Teso Diego~Calanzone and Antonio Vergari.
\newblock Logically consistent language models via neuro-symbolic integration.
\newblock {\em arXiv}, 2409.13724, September 2024.

\bibitem{10323951}
Wenji Fang, Yao Lu, Shang Liu, Qijun Zhang, Ceyu Xu, Lisa~Wu Wills, Hongce Zhang, and Zhiyao Xie.
\newblock Masterrtl: A pre-synthesis ppa estimation framework for any rtl design.
\newblock In {\em 2023 IEEE/ACM International Conference on Computer Aided Design (ICCAD)}, pages 1--9, 2023.

\bibitem{gupta1997path}
Rajiv Gupta, DA~Benson, and Jesse~Zhixi Fang.
\newblock Path profile guided partial dead code elimination using predication.
\newblock In {\em Proceedings 1997 International Conference on Parallel Architectures and Compilation Techniques}, pages 102--113. IEEE, 1997.

\bibitem{hopcroft1969formal}
John~E Hopcroft and Jeffrey~D Ullman.
\newblock {\em Formal languages and their relation to automata}.
\newblock Addison-Wesley Longman Publishing Co., Inc., 1969.

\bibitem{hu2023codepromptingneuralsymbolic}
Yi~Hu, Haotong Yang, Zhouchen Lin, and Muhan Zhang.
\newblock Code prompting: a neural symbolic method for complex reasoning in large language models, 2023.

\bibitem{knoop1994partial}
Jens Knoop, Oliver R{\"u}thing, and Bernhard Steffen.
\newblock Partial dead code elimination.
\newblock {\em ACM Sigplan Notices}, 29(6):147--158, 1994.

\bibitem{laforest10}
Charles~Eric LaForest and J.~Gregory Steffan.
\newblock Efficient multi-ported memories for fpgas.
\newblock In {\em Proceedings of the 18th Annual ACM/SIGDA International Symposium on Field Programmable Gate Arrays (FPGA '10)}, pages 41--50, Monterey, CA, USA, 2010. ACM.

\bibitem{langley00}
P.~Langley.
\newblock Crafting papers on machine learning.
\newblock In Pat Langley, editor, {\em Proceedings of the 17th International Conference on Machine Learning (ICML 2000)}, pages 1207--1216, Stanford, CA, 2000. Morgan Kaufmann.

\bibitem{liu:2023:chipnemo}
Mingjie Liu, Teodor{-}Dumitru Ene, Robert Kirby, Chris Cheng, Nathaniel~Ross Pinckney, Rongjian Liang, Jonah Alben, Himyanshu Anand, Sanmitra Banerjee, Ismet Bayraktaroglu, Bonita Bhaskaran, Bryan Catanzaro, Arjun Chaudhuri, Sharon Clay, Bill Dally, Laura Dang, Parikshit Deshpande, Siddhanth Dhodhi, Sameer Halepete, Eric Hill, Jiashang Hu, Sumit Jain, Brucek Khailany, Kishor Kunal, Xiaowei Li, Hao Liu, Stuart~F. Oberman, Sujeet Omar, Sreedhar Pratty, Jonathan Raiman, Ambar Sarkar, Zhengjiang Shao, Hanfei Sun, Pratik~P. Suthar, Varun Tej, Kaizhe Xu, and Haoxing Ren.
\newblock Chipnemo: Domain-adapted llms for chip design.
\newblock {\em CoRR}, abs/2311.00176, 2023.

\bibitem{liu:2023:verilogeval}
Mingjie Liu, Nathaniel~Ross Pinckney, Brucek Khailany, and Haoxing Ren.
\newblock Invited paper: Verilogeval: Evaluating large language models for verilog code generation.
\newblock In {\em {IEEE/ACM} International Conference on Computer Aided Design, {ICCAD}}. {IEEE}, 2023.

\bibitem{liu:2024:rtlcoder}
Shang Liu, Wenji Fang, Yao Lu, Jing Wang, Qijun Zhang, Hongce Zhang, and Zhiyao Xie.
\newblock Rtlcoder: Fully open-source and efficient llm-assisted rtl code generation technique.
\newblock {\em IEEE Transactions on Computer-Aided Design of Integrated Circuits and Systems}, 2024.

\bibitem{ma2020hypervisor}
Jiacheng Ma, Gefei Zuo, Kevin Loughlin, Xiaohe Cheng, Yanqiang Liu, Abel~Mulugeta Eneyew, Zhengwei Qi, and Baris Kasikci.
\newblock A hypervisor for shared-memory fpga platforms.
\newblock In {\em Proceedings of the Twenty-Fifth International Conference on Architectural Support for Programming Languages and Operating Systems}, pages 827--844, 2020.

\bibitem{moore1956gedanken}
Edward~F Moore et~al.
\newblock Gedanken-experiments on sequential machines.
\newblock {\em Automata studies}, 34:129--153, 1956.

\bibitem{openai2023gpt4}
OpenAI, Josh Achiam, Steven Adler, Sandhini Agarwal, , and many others.
\newblock Gpt-4 technical report.
\newblock {\em arXiv}, 2303.08774, March 2023.

\bibitem{palnitkar2003verilog}
Samir Palnitkar.
\newblock {\em Verilog HDL: a guide to digital design and synthesis}, volume~1.
\newblock Prentice Hall Professional, 2003.

\bibitem{subexpression}
R.~Pasko, P.~Schaumont, V.~Derudder, S.~Vernalde, and Daniela Duracková.
\newblock A new algorithm for elimination of common subexpressions.
\newblock {\em Computer-Aided Design of Integrated Circuits and Systems, IEEE Transactions on}, 18:58 -- 68, 02 1999.

\bibitem{schultz2023optimizing}
Jim Schultz.
\newblock Optimizing the rtl design flow with real-time ppa analysis.
\newblock {\em Synopsys Silicon to Systems Blog}, N/A, March 2023.
\newblock Blog Post. URL: https://www.synopsys.com/blogs/silicon-to-systems/optimizing-rtl-design-flow-real-time-ppa-analysis/ (Accessed January 26, 2025).

\bibitem{synopsysDCUltra}
Synopsys.
\newblock Dc ultra for synthesis and test.
\newblock \url{https://www.synopsys.com/implementation-and-signoff/rtl-synthesis-test/dc-ultra.html}.

\bibitem{Takamaeda:2015:ARC:Pyverilog}
Shinya Takamaeda-Yamazaki.
\newblock Pyverilog: A python-based hardware design processing toolkit for verilog hdl.
\newblock In {\em Applied Reconfigurable Computing}, volume 9040 of {\em Lecture Notes in Computer Science}, pages 451--460. Springer International Publishing, Apr 2015.

\bibitem{taraate2022digital}
Vaibbhav Taraate.
\newblock {\em Digital logic design using verilog}.
\newblock Springer, 2022.

\bibitem{thakur:2024:verigen}
Shailja Thakur, Baleegh Ahmad, Hammond Pearce, Benjamin Tan, Brendan Dolan{-}Gavitt, Ramesh Karri, and Siddharth Garg.
\newblock Verigen: {A} large language model for verilog code generation.
\newblock {\em {ACM} Trans. Design Autom. Electr. Syst.}, 2024.

\bibitem{tsai:2024:rtlfixer}
Yunda Tsai, Mingjie Liu, and Haoxing Ren.
\newblock Rtlfixer: Automatically fixing {RTL} syntax errors with large language model.
\newblock In {\em Proceedings of the 61st {ACM/IEEE} Design Automation Conference, {DAC}}. {ACM}, 2024.

\bibitem{vahid2010digital}
Frank Vahid.
\newblock Digital design with rtl design, vhdl, and verilog (2nd edition).
\newblock {\em John Wiley \& Sons Inc}, 2, January 2010.
\newblock ISBN-13: 978-0470531082, 575 pages.

\bibitem{wan2024neurosymbolic}
Zishen Wan, Che-Kai Liu, Hanchen Yang, Ritik Raj, Chaojian Li, Haoran You, Yonggan Fu, Cheng Wan, Sixu Li, Youbin Kim, Ananda Samajdar, Yingyan~Celine Lin, Mohamed Ibrahim, Jan~M. Rabaey, Tushar Krishna, and Arijit Raychowdhury.
\newblock Cross-layer design for neuro-symbolic ai: From workload characterization to hardware acceleration.
\newblock {\em arXiv}, 2409.13153, September 2024.
\newblock Available at https://arxiv.org/abs/2409.13153.

\bibitem{wan2024neurosymbolicTowards}
Zishen Wan, Che-Kai Liu, Hanchen Yang, Ritik Raj, Chaojian Li, Haoran You, Yonggan Fu, Cheng Wan, Sixu Li, Youbin Kim, Ananda Samajdar, Yingyan~Celine Lin, Mohamed Ibrahim, Jan~M. Rabaey, Tushar Krishna, and Arijit Raychowdhury.
\newblock Towards efficient neuro-symbolic ai: From workload characterization to hardware architecture.
\newblock {\em arXiv}, 2409.13153, September 2024.
\newblock Published in IEEE Transactions on Circuits and Systems for Artificial Intelligence (TCASAI), 2024.

\bibitem{wang2009electronic}
Laung-Terng Wang, Yao-Wen Chang, and Kwang-Ting~Tim Cheng.
\newblock {\em Electronic design automation: synthesis, verification, and test}.
\newblock Morgan Kaufmann, 2009.

\bibitem{wang2023muxoptimization}
Zicheng Wang, Hailong You, Jie Wang, Meihua Liu, Yu~Su, and Yong Zhang.
\newblock Optimization of multiplexer combination in rtl logic synthesis.
\newblock In {\em Proceedings of the 2023 International Symposium of Electronics Design Automation (ISEDA)}, page N/A. IEEE, May 2023.

\bibitem{wolf2013yosys}
Clifford Wolf, Johann Glaser, and Johannes Kepler.
\newblock Yosys-a free verilog synthesis suite.
\newblock In {\em Proceedings of the Forum on Specification and Design Languages (FDL)}, 2013.
\newblock Yosys is the first open-source Verilog synthesis suite supporting a wide range of synthesizable Verilog features.

\bibitem{yang2023neurosymbolic}
Sen Yang, Xin Li, Leyang Cui, Lidong Bing, and Wai Lam.
\newblock Neuro-symbolic integration brings causal and reliable reasoning proofs.
\newblock {\em arXiv}, 2311.09802, November 2023.
\newblock Code available at this URL.

\bibitem{rtlrewriter}
Xufeng Yao, Yiwen Wang, Xing Li, Yingzhao Lian, Chen Ran, Lei Chen, Mingxuan Yuan, Hong Xu, and Bei Yu.
\newblock Rtlrewriter: Methodologies for large models aided rtl code optimization, 09 2024.

\bibitem{ye2023comprehensivecapabilityanalysisgpt3}
Junjie Ye, Xuanting Chen, Nuo Xu, Can Zu, Zekai Shao, Shichun Liu, Yuhan Cui, Zeyang Zhou, Chao Gong, Yang Shen, Jie Zhou, Siming Chen, Tao Gui, Qi~Zhang, and Xuanjing Huang.
\newblock A comprehensive capability analysis of gpt-3 and gpt-3.5 series models, 2023.

\bibitem{994596}
Haifeng Zhou, Zhenghui Lin, and Wei Cao.
\newblock Research on vhdl rtl synthesis system.
\newblock In {\em Proceedings First IEEE International Workshop on Electronic Design, Test and Applications '2002}, pages 99--103, 2002.

\end{thebibliography}
